\documentstyle[psfig]{l-aa}
\def\Xray{\hbox{X-ray}}
\def\Msun{\hbox{M${}_\odot$}}
\def\Rsun{\hbox{R${}_\odot$}}

\def\CygX{\hbox{Cyg X-}}
\def\CygnusX{\hbox{Cygnus X-}}
\def\micron{\hbox{$\mu$m}}

\def\HIBrg{\hbox{H\,{\sc i}\,Br$\gamma$}}
\def\HeI{\hbox{He\,{\sc i}}}
\def\HeII{\hbox{He\,{\sc ii}}}
\def\HeIII{\hbox{He\,{\sc iii}}}

\def\NIII{\hbox{N\,{\sc iii}}}
\def\NIV{\hbox{N\,{\sc iv}}}
\def\NV{\hbox{N\,{\sc v}}}

\def\CaII{\hbox{Ca\,{\sc ii}}}
\def\water{\hbox{H$_2$O}}
\def\cotwo{\hbox{CO$_2$}}
\def\kms{\hbox{km\,s$^{-1}$}}
\def\nodata{\o}
\def\notdet{$\cdots$}
\def\doublebox!#1!#2!{\hbox{$\vcenter{
  \tabskip=0em\halign{\strut##\hfil\cr #1\cr #2\cr}}$}}
\def\pho{\hbox{$\phantom{0}$}}
\def\phoo{\hbox{$\phantom{00}$}}
\def\phC{\hbox{$\phantom{100}$}}
\def\phM{\hbox{$\phantom{1000}$}}
\def\th{\hbox{$^{\rm th}$}}
\begin{document}

\thesaurus{05(08.02.1; 08.09.2 Cygnus X-3; 08.23.2; 13.25.5)}

\title{The Wolf-Rayet counterpart of Cygnus~X-3\thanks{Based on
observations made at the United Kingdom Infrared Telescope on Mauna
Kea, operated on the island of Hawaii by the Royal Observatories, and
on observations made at the William Herschel Telescope, operated on
the island of La Palma by the Royal Greenwich Observatory in the
Spanish Observatorio del Roque de los Muchachos of the Instituto de
Astrof\'\i{}sica de Canarias}}

\author{M.H.~van~Kerkwijk\inst{1,2}
\and    T.R.~Geballe\inst{3}
\and    D.L.~King\inst{4}
\and    M.~van der Klis\inst{1}
\and    J.~van~Paradijs\inst{1,5}
}
\institute{%
     Astronomical Institute ``Anton Pannekoek'', University of Amsterdam,
     and Center for High-Energy Astrophysics (CHEAF), Kruislaan 403, 
     1098 SJ\ \ Amsterdam, The Netherlands
\and Department of Astronomy, California Institute of Technology, m.s.~105-24,
     Pasadena, CA~91125, USA
\and Joint Astronomy Centre, 665 Komohana Street, Hilo, HI~96720, USA
\and Royal Greenwich Observatory, Madingley Road, Cambridge CB3 0EZ, UK
\and Physics Department, University of Alabama in Huntsville, 
     Huntsville, AL~35899, USA
}
\date{Submitted 1995 September 9}
\offprints{M.H.\ van Kerkwijk (Caltech)}
\maketitle

\begin{abstract}

We present orbital-phase resolved I and K-band spectroscopy of
Cygnus~X-3.  All spectra show emission lines characteristic of
Wolf-Rayet stars of the WN subclass.  On time scales longer than about
one day, the line strengths show large changes, both in flux and in
equivalent width.  In addition, the line ratios change, corresponding
to a variation in spectral subtype of WN6/7 to~WN4/5.  We confirm the
finding that at times when the emission lines are weak, they shift in
wavelength as a function of orbital phase, with maximum blueshift
coinciding with infrared and \Xray\ minimum, and maximum redshift half
an orbit later.  Furthermore, we confirm the prediction -- made on the
basis of previous observations -- that at times when the emission
lines are strong, no clear wavelength shifts are observed.  We
describe a simplified, but detailed model for the system, in which the
companion of the \Xray\ source is a Wolf-Rayet star whose wind is at
times ionised by the \Xray\ source, except for the part in the star's
shadow.  With this model, the observed spectral variations can be
reproduced with only a small number of free parameters.  We discuss
and verify the ramifications of this model, and find that, in general,
the observed properties can be understood.  We conclude that \CygX3 is
a Wolf-Rayet/\Xray\ binary.

\end{abstract}

\section{Introduction\label{sec:intro}}

Cygnus X-3 is a bright \Xray\ source that is peculiar among \Xray\
binaries.  It has huge radio outbursts, mildly relativistic
($\beta\simeq0.3$) jets, smooth 4.8\,hour orbital modulation of its
\Xray\ light curve, and a rapid increase of the orbital period on a
time scale of 850\,000 years.  It also has a very strong iron line in
its \Xray\ spectrum, is very bright in the infrared, and has been
claimed to be detected at very high energies (for reviews of its
properties, see, e.g., Bonnet-Bidaud \& Chardin \cite{bonnc:88}; Van
der Klis \cite{vdkl:93}).

One of the first models for \CygX3 was put forward by Van den Heuvel
\& De Loore (\cite{vdhedl:73}).  They suggested that the system is
composed of a compact object and a helium star of several solar
masses, and that it represents a late evolutionary stage of massive
\Xray\ binaries (the so-called `second Wolf-Rayet phase'; Van den
Heuvel \cite{vdhe:76}).  Massive helium stars have been identified
observationally with the group of `classical', or population~I
Wolf-Rayet stars (Van der Hucht et al.\ \cite{vdhu&a:81}).  Such stars
have strong winds, which in \CygX3 would be the underlying cause for
the \Xray\ modulation (due to scattering of the X~rays), the increase
of the orbital period (due to the loss of angular momentum) and the
brightness in the infrared (due to free-free emission in the wind).
In the course of further evolution, the helium star is likely to
explode as a supernova.  If the system is not disrupted, a binary such
as the Hulse-Taylor pulsar PSR\,1913+16 could be formed (Flannery \&
Van den Heuvel \cite{flanvdh:75}).

In this model, it is predicted that the optical/infrared counterpart
shows Wolf-Rayet features in its spectrum.  This prediction was
confirmed by Van Kerkwijk et al.\ (\cite{vker&a:92}, hereafter
Paper~I), who found strong, broad emission lines of \HeI\ and \HeII\
-- but no evidence for hydrogen -- in I and K-band spectra of \CygX3,
as expected for a Wolf-Rayet star of spectral type WN7.  In subsequent
observations, it was found (Van Kerkwijk \cite{vker:93a}, hereafter
Paper~II) that large changes in the absolute and relative strengths of
the emission lines had occurred.  Furthermore, the spectra showed
orbital-phase dependent wavelength shifts of the emission lines, with
maximum blueshift occurring at the time of infrared and \Xray\
minimum, and maximum redshift half an orbit later.

It was shown that these wavelength shifts could be understood if the
Wolf-Rayet wind were almost completely ionised by the \Xray\ source at
the time of the observations, except in the part shadowed by the
helium star.  It was found that both the wavelength shifts and the
modulation of the infrared continuum could be reproduced with a
detailed model (with only a small number of free parameters).  Based
on the model, it was predicted that at times when strong emission
lines were present in the infrared spectra, there would be little
modulation of the lines and the continuum, whereas at times when the
emission lines were weak, there would be a clear modulation of the
lines and continuum.  For the latter case, it was expected that at
high resolution the line profile would be resolved in two components.

Based on the idea that a stronger \Xray\ source would ionise more of
the wind, it was also predicted that the \Xray\ source should be in
its low state (low flux, hard spectrum) when the emission lines were
strong, and in its high state (high flux, soft spectrum) when they
were weak.  Kitamoto et al.\ (\cite{kita&a:94}), however, found that
\Xray\ and radio data indicated that the source was in its high state
on 1991 June 21, when the infrared lines were strong, while radio data
indicated it was in its low state on 1992 May 29, when the lines were
weak.  They suggested a modification of the model presented in
Paper~II, in which the source's state at all wavelengths was a
function of the mass-loss rate of the Wolf-Rayet star only.  In this
case, a high mass-loss rate would lead to increased infrared and
\Xray\ fluxes, as well as to radio outbursts.  At the same time, the
wind would become optically thick to X~rays, leading to a lower degree
of ionisation, and hence stronger infrared emission lines.

In this paper, we discuss the model for the infrared continuum and
lines in detail, and compare it to the available observations.  In
Sect.~\ref{sec:obsred}, the procedures used for making and reducing
the observations are described, both for the observations presented in
Papers~I and~II, and for a number of additional observations.  We
present the observations in Sect.~\ref{sec:spectra}, and point out the
characteristic similarities and differences shown by the spectra.  In
Sect.~\ref{sec:model}, we describe our model for the system, and use
it to calculate light curves and line profiles as a function of
orbital phase.  Furthermore, we qualitatively interpret the long-term
changes, and verify the predictions for the lines made in Paper~II.
In Sect.~\ref{sec:massloss}, we estimate the velocity in the wind, and
discuss different estimates of the mass-loss rate.  We discuss the
ramifications expected for a more realistic treatment of the wind in
Sect.~\ref{sec:ramifications}.  In Sect.~\ref{sec:fcont}, we estimate
the infrared flux distribution of \CygX3, and compare it with the one
predicted for our model, and with the ones observed for other
Wolf-Rayet stars.  We draw conclusions about the nature of \CygX3 in
Sect.~\ref{sec:bigdisc}.

\section{Observations and data reduction\label{sec:obsred}} 

The present study is based on 37 I-band and 16 K-band spectra,
obtained in 1991, 1992 and 1993.  A log of the observations is given
in Tables \ref{tab:ilog} and \ref{tab:klog}.  Below, we discuss the
procedures used for making and reducing the observations.

\begin{table}[t]
\caption[]{The I-band observations\label{tab:ilog}}
\hbox to\hsize{%
\begin{tabular}{lll@{}}
\hline
JD$_{\rm bar., mid-exp.}$& t$_{\rm exp}$& $\phi_{\rm X}{}^{\rm a}$\\
($-$2440000)&         (min.)\\
\hline
\multicolumn{3}{@{}l@{}}{\em 1991 June 21, 0.72--1.05\,$\mu$m}\\
8428.691& 30& 0.67\\
\phM.712& 27& 0.77\\[1mm]
\multicolumn{3}{@{}l@{}}{\em 1992 July 25--27, 0.85--1.10\,$\mu$m}\\
8828.465& 40& 0.65\\
\phM.499& 40& 0.82\\
\phM.530& 40& 0.97\\
\phM.563& 40& 0.14\\
\phM.595& 40& 0.30\\
\phM.626& 40& 0.45\\
\phM.657& 40& 0.61\\
\phM.688& 40& 0.76\\[1mm]
8829.470& 40& 0.68\\
\phM.502& 40& 0.84\\
\phM.533& 40& 0.00\\
\phM.563& 40& 0.15\\
\phM.595& 40& 0.31\\
\phM.625& 40& 0.46\\
\phM.656& 40& 0.61\\
\phM.686& 40& 0.76\\
\hline
\end{tabular}
\hfill
\begin{tabular}{lll@{}}
\hline
JD$_{\rm bar., mid-exp.}$& t$_{\rm exp}$& $\phi_{\rm X}{}^{\rm a}$\\
($-$2440000)&         (min.)\\
\hline
8830.448& 40& 0.58\\
\phM.478& 40& 0.73\\
\phM.508& 40& 0.88\\
\phM.538& 40& 0.03\\
\phM.572& 40& 0.20\\
\phM.602& 40& 0.35\\
\phM.632& 40& 0.50\\
\phM.662& 40& 0.65\\[1mm]
\multicolumn{3}{@{}l@{}}{\em 1993 June 13, 14, 0.85--1.10\,$\mu$m}\\
9151.581& 40& 0.74\\
\phM.614& 40& 0.90\\
\phM.648& 40& 0.08\\
\phM.676& 40& 0.22\\
\phM.708& 40& 0.38\\[1mm]
9152.562& 40& 0.65\\
\phM.592& 40& 0.80\\
\phM.622& 40& 0.95\\
\phM.652& 40& 0.10\\
\phM.684& 40& 0.26\\
\phM.710& 33& 0.39\\
\hline
\end{tabular}
}
\vskip1mm\noindent
$^{\rm a}$ Using the quadratic ephemeris of Kitamoto et al.\
\cite{kita&a:95}: $T_0={\rm{}JD_{bar}}\;2440949.8923(7)$, 
$P_0=0.19968430(9)\,$d, $\dot{P}=6.48(24)10^{-10}$
\end{table}

\subsection{The I-band observations\label{sec:obsi}}

The I-band spectra were obtained using the ISIS spectrograph at the
Cassegrain focus of the William Herschel Telescope (WHT), Observatorio
de Roque de las Muchachas, La Palma.  The $158\,{\rm{}l\,mm^{-1}}$
grating, blazed at 6500\AA, was used in first order.  The central
wavelength was set to 0.8730\,\micron\ in 1991 (Paper~I), and -- in
view of the low signal at shorter wavelengths -- to 1.01\micron\ in
1992 and 1993.  An OG530 (1991) or RG630 (1992, 1993) filter was used
to block the higher orders.  The slit width was set to $2\arcsec$
(except for the first observation in 1992, when a slit of $1\farcs5$
was used).  The detector was a cooled EEV CCD with 1242$\times$1152
square pixels of $22.5\,\micron$ on the side, corresponding to
$0\farcs33$ on the sky.  With this setup, the dispersion is
$\sim\!2.7\,\AA\,{\rm{}pix^{-1}}$, the resolution $\sim\!16\,\AA$ and
the wavelength coverage $\sim\!0.34\,\micron$.  In 1992, the chip
temperature was raised in an attempt to increase the near-infrared
response.  The slight increase in the sensitivity, however, was offset
by the increased noise due to the dark current.  

At 20\th\ magnitude in I, \CygX3 was not visible on the slit-viewing
television.  In order to ensure that it was observed, in 1991 the slit
was placed over the stars called A and C on the I-band finding chart
of Wagner et al.\ (\cite{wagn&a:89}).  J, H and K-band images (Joyce
\cite{joyc:90}), however, revealed that \CygX3 is actually about
$1\farcs2$ SE of the line joining stars A and C.  In the 1992 and 1993
runs, therefore, the slit was set accurately over \CygX3 by first
placing it over stars~A and C, and then rotating it anticlockwise by
$2\fdg3$, keeping star~A in the slit.  This has the advantage that
star~A can still be used to correct for telluric absorption.  In 1992
and 1993, spectra of several Wolf-Rayet stars and low-mass \Xray\
binaries were taken for comparison with the spectra of \CygX3.  In all
runs, the observing conditions were good.

The spectra were reduced using the MIDAS reduction package and
additional routines running in the MIDAS environment.  The usual steps
of bias subtraction, flat-field correction, sky subtraction, optimal
extraction (Horne \cite{horn:86}), cosmic-ray removal and wavelength
calibration were carried out (for details, see Van Kerkwijk
\cite{vker:93b}).  

The noisy 1992 data, however, caused some difficulty during the
extraction process.  Horne (private communication) suggested using the
spatial profile of star~A to optimally extract the spectrum of \CygX3.
We verified this using the 1991 and 1993 spectra, and found that the
results were virtually indistinguishable from those obtained using the
spatial profile determined from \CygX3 itself.  Indeed, the stronger
signal of star~A longward of $1.05\,\micron$ led to a better
extraction at those wavelengths.  We therefore used the spatial
profile of star~A for the extraction of all spectra.

\begin{figure}[tb]
\centerline{\hbox{\psfig{figure=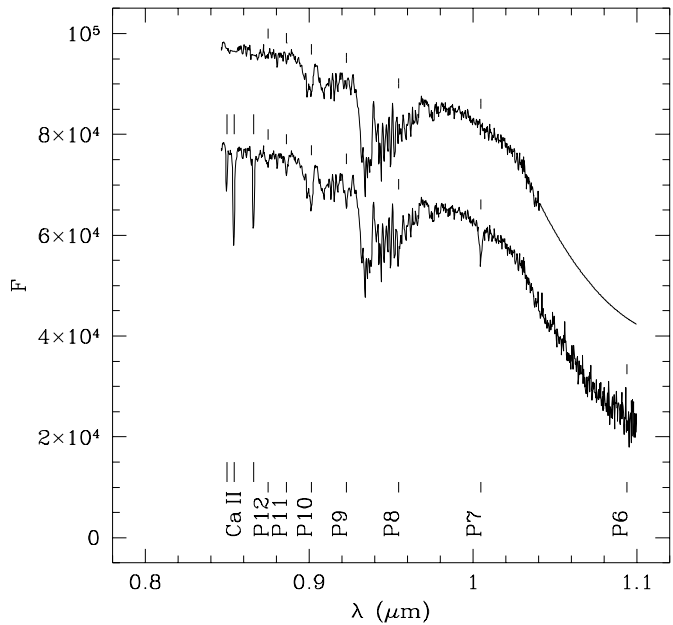,bbllx=32pt,bburx=322pt,width=8.8cm}}}
\caption[]{The I-band spectrum of star~A.  The lower curve shows the
average of the spectra observed on 1993 May 13.  The
identifications of the stellar features -- members of the Paschen
series and the \CaII\ triplet -- are indicated.  The upper curve shows
the spectrum after the removal of these features and the substitution
of the noisy long-wavelength part by a polynomial (for details, see
text).  With such `cleaned' spectra, the spectra of Cyg~X-3 have been
corrected for the telluric water-vapour absorption
lines\label{fig:stara}}
\end{figure}

Each extracted spectrum was corrected for telluric water vapour
absorption by dividing it by the spectrum of star~A taken in the same
exposure.  For this purpose, the strong stellar features of star~A --
the \CaII\ triplet and Paschen lines P6--P16 -- were removed (the
relative strengths of the lines indicate spectral type F5--G0; for
such a spectral type no other strong features are expected).
Furthermore, for the spectra of star~A obtained in 1992 and 1993, the
long-wavelength part -- which is dominated by Poisson noise rather
than telluric features -- was replaced by a best-fitting polynomial.
For the individual spectra, the range with $\lambda\ga1.01\,\micron$
was replaced, and for the average spectra, the range with
$\lambda\ga1.04\,\micron$.  After this substitution, the spectra were
inspected by eye, and (slight) discontinuities between the substituted
and non-substituted part were corrected for by smoothing the spectra
locally.  An example is shown in Fig.~\ref{fig:stara}.

As a final step, the \CygX3 spectra were binned in 3-pixel wide bins
($\sim\!8\,\AA$, about half a resolution element).  Per bin, the
signal-to-noise ratio is about 10 at $1\,\micron$ for the spectra
taken in 1991 and 1993, while it is about 6 for those taken in 1992
(due to the lower flux level of the source).

\begin{table}[t]
\caption[]{The K-band observations\label{tab:klog}}
\begin{tabular}{llll@{}}
\hline
JD$_{\rm bar., mid-exp.}$& t$_{\rm obs}{}^{\rm a}$& t$_{\rm exp}{}^{\rm b}$& 
$\phi_{\rm X}{}^{\rm c}$\\
($-$2440000)&         (min.)& (sec.)\\
\hline
\multicolumn{3}{l@{}}{\em 1991 June 29, 2.0--2.4\,$\mu$m}\\
 8437.106& 20& 16$\times$6$\times$5&  0.81\\[1mm]
\multicolumn{3}{l@{}}{\em 1991 July 20, 2.0--2.4\,$\mu$m}\\
 8458.102& 27& 16$\times$6$\times$10& 0.95\\[1mm]
\multicolumn{3}{l@{}}{\em 1992 May 29, 2.0--2.4\,$\mu$m}\\
 8771.930& 22& 12$\times$6$\times$10& 0.53\\
\phM .951& 20& 12$\times$6$\times$10& 0.64\\
\phM .977& 20& 12$\times$6$\times$10& 0.77\\
\phC2.008& 20& 12$\times$6$\times$10& 0.92\\
\phM .023& 20& 12$\times$6$\times$10& 0.00\\
\phM .053& 30& 18$\times$6$\times$10& 0.15\\
\phM .090& 20& 12$\times$6$\times$10& 0.33\\
\phM .105& 20& 12$\times$6$\times$10& 0.41\\[1mm]
\multicolumn{3}{l@{}}{\em 1992 July 24, 2.03--2.23\,$\mu$m}\\
 8828.033& 30& 20$\times$6$\times$10& 0.48\\[1mm]
\multicolumn{3}{l@{}}{\em 1993 July 15, 2.03--2.23\,$\mu$m}\\
 9183.931& 30& 20$\times$6$\times$10& 0.74\\
\phM .965& 30& 20$\times$6$\times$10& 0.91\\
\phM .987& 32& 22$\times$6$\times$10& 0.02\\
\phC4.018& 21& 16$\times$6$\times$10& 0.18\\
\phM .075& 30& 20$\times$6$\times$10& 0.46\\
\hline
\end{tabular}
\vskip1mm\noindent
$^{\rm a}$ Time spent on source\\
$^{\rm b}$ Number of integrations times the number of exposures per
integration times the exposure time of one exposure (see
Sect.~\ref{sec:obsk})\\
$^{\rm c}$ Using the ephemeris of Kitamoto et al.\ \cite{kita&a:95}
(Table~\ref{tab:ilog}, note a)
\end{table}

\subsection{The K-band observations\label{sec:obsk}} 

The K-band spectra were obtained at the United Kingdom Infrared
Telescope (UKIRT) on Mauna Kea, Hawaii, using the Cooled Grating
Spectrometer CGS4.  On 1991 June 29, a first Service time
spectrum was taken in good weather conditions (Paper~I).  The 75\,l/mm
grating was used in first order, combined with a filter to block the
higher orders.  The detector was a SBRC InSb array with 58$\times$62
elements.  The 150\,mm focal-length camera was used, for which the
pixel size corresponds to $3\farcs1$ on the sky.  The slit width was
chosen to match the pixel size.  The wavelength range covered with
this setup is 2.0--2.4\,\micron, at 0.0063\,\micron/pixel.

A second Service observation was obtained on 1991 July 20 under
non-photometric conditions, using the same setup.  On 1992 May 29,
\CygX3 was observed for almost one complete orbital period (Paper~II).
Spectra of several Wolf-Rayet stars and low-mass \Xray\ binaries were
taken for comparison that night and the next.  During both nights the
observing conditions were good.  On 1992 July 24, another Service
observation was taken under non-photometric conditions.  At that time,
the optics of the camera had been changed, so that the projected pixel
size on the sky was reduced to $1\farcs5$ (the slit width was set
accordingly), and the wavelength range to 2.03--2.23\,\micron, at
0.0032\,\micron/pixel.  The same setup was used on 1993 July 15, to
observe \CygX3 again for nearly a full orbital period.  The weather
conditions during this night were not photometric.

In the K band, \CygX3 is a 12\th\ magnitude source, and thus, unlike
in the I band, it could be centred on the slit without problem, by
peaking up in a given row of pixels.  In all runs, an observation
consisted of the sum of 6 to 11 pairs of integrations, taken at two
different positions on the chip.  The individual integrations are
composed of 6 exposures, taken at detector positions offset by
multiples of one third of a pixel, so that one obtains an effective
resolution determined by the size of one pixel.  The reduction process
consisted of the following steps (for details, see Van Kerkwijk
\cite{vker:93b}): (i) bias and flat-field correction; (ii) combination
of exposures into integrations, integrations into pairs, and pairs of
integrations into observations; (iii) extraction of the spectra; (iv)
ripple correction; (v) wavelength calibration; and (vi) flux
calibration.  Of these, steps (i) and (ii) are performed on-line
during the night.  For the extraction, we used the optimal extraction
algorithm developed by Horne (\cite{horn:86}) if more than one row of
pixels was illuminated, which was the case for all runs except that of
1992 May (for the 1991 June spectra, the illumination of the rows for
\CygX3 and the standard were slightly different; hence, the slope of
the continuum of the reduced spectrum presented here is slightly
different from that presented in Paper~I, for which the spectra were
extracted using only the row that contained most of the light).

In all runs, spectra of bright stars of spectral types F and A, taken
interspersed with the observations of \CygX3, were used for the
correction for telluric water-vapour and carbon-dioxide absorption
features, and for flux calibration.  In spectra of stars of these
types, the only strong stellar feature is \HIBrg.  For the
calibration, this feature was removed.  The stars used were HR\,7796
(F8I, $K=0.72$) in 1991, 1992 July and 1993, and HR\,8028 (A1V,
$K=3.80$) in 1992 May.  In 1993, we also used HR\,7847 (F5I, $K=3.51$)
to verify the calibration.  The correction for telluric absorption
features proved to be highly satisfactory for almost all spectra.  The
only exceptions are found among some of the 1992 May 29 spectra, which
were taken at rapidly decreasing airmass (most notably the first).
The flux calibration is reliable (better than $\sim\!5$\% in the
absolute level) only for the spectra taken in good conditions, i.e.,
those taken on 1991 June 29 and 1992 May 29.  For the other runs, we
expect, from a comparison of different observations of the flux
standards, that the absolute fluxes are accurate to $\sim\!20$\%.

\section{The spectra\label{sec:spectra}}

\subsection{The I-band spectra}

\begin{figure}[tb]
\centerline{\hbox{\psfig{figure=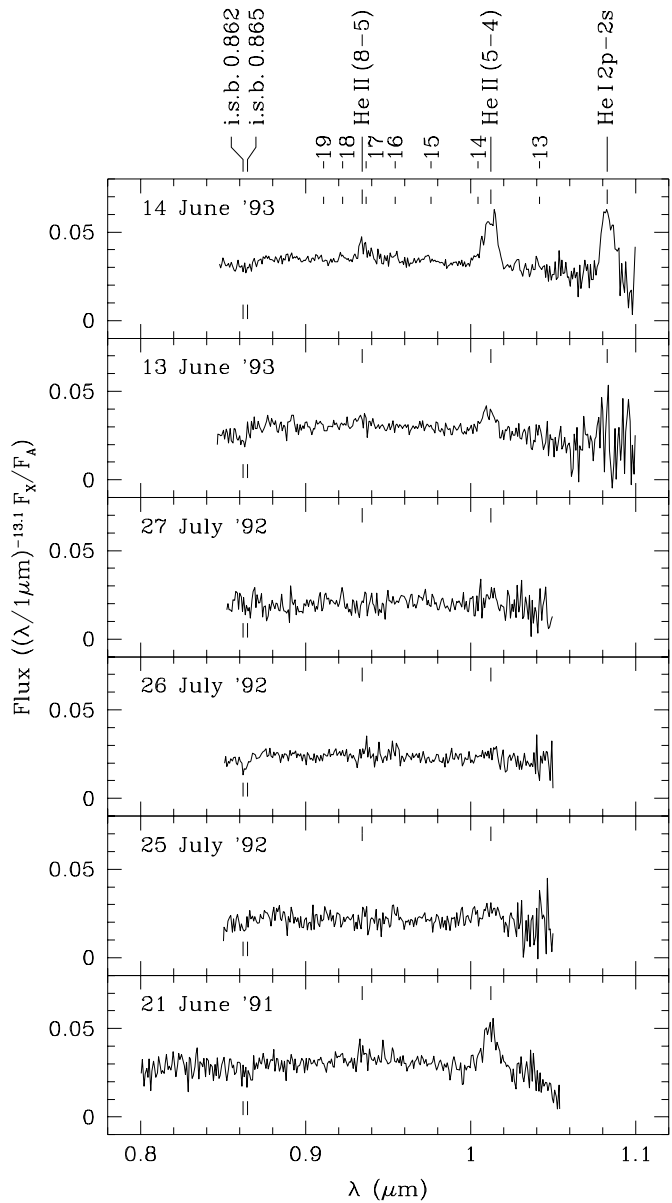,bbllx=32pt,bburx=322pt,width=8.8cm}}}
\caption[ ]{The average I-band spectra of \CygX3.  The spectra shown
are corrected for telluric features by dividing them by spectra of
star~A from which the stellar features are removed (see text,
Fig.~\ref{fig:stara}).  For clarity, the ratio spectra were multiplied
with $(\lambda/1\micron)^{-13.1}$, and for the 1992 spectra the noisy
long-wavelength portion was omitted.  At the top of the figure, the
identifications of the strongest lines are indicated.  Also indicated
are some weaker lines that seem to be present in the 1993 June 13
spectrum.  For these, only a number is given, which refers to the
upper level of \HeII\,$(n-6)$\label{fig:iav}}
\end{figure}

It was found in Paper~I that the continuum of the 1991 I-band spectrum
could be well represented by a power law of the form
$F_X/F_A=C\lambda^\beta$.  This was confirmed for the 1992 and 1993
spectra: although the best-fitting constant of proportionality $C$
varies, the power-law index $\beta$ is very similar, with an average
value of 13.1 (in the range 0.85 to 1.0\,\micron, but excluding
\HeII\,$(8-5)$).  In the figures, the spectra are flattened by
divideding them by $\lambda^{13.1}$ ($\lambda$ in \micron).  The
averages of spectra taken in one night are shown in
Fig.~\ref{fig:iav}.  Identifications of the lines are shown in the
figure and listed in Table~\ref{tab:iid}.

\paragraph{1991 June 21.} 
In this spectrum, presented also in Paper~I, the most prominent
emission line is of \HeII\,$(5-4)$ at 1.0123\micron.  Weak emission
seems to be present as well in \HeII\,$(8-5)$, at 0.9345\,\micron.
Furthermore, the spectrum shows an absorption feature at
0.864\,\micron, the reality of which is confirmed by its presence in
the other I-band spectra.  This feature is due to the interstellar
bands at 0.8620 and 0.8649\,\micron\ (Herbig \& Leka \cite{herbl:91}).
Note that the flux relative to star~A is probably underestimated for
this spectrum, since the source was observed offset from the centre of
the slit (see Sect.~\ref{sec:obsi}).

\begin{figure}[tb]
\centerline{\hbox{\psfig{figure=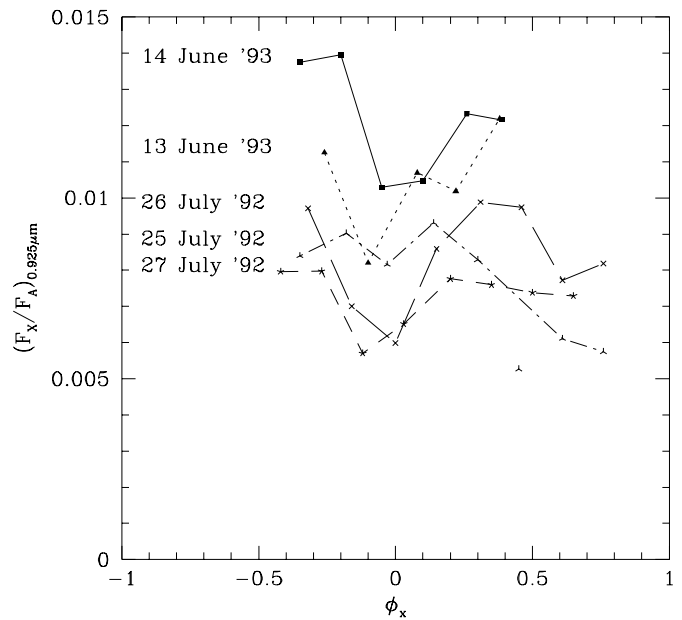,bbllx=32pt,bburx=322pt,width=8.8cm}}}
\caption[]{The 0.925\,\micron\ continuum flux of \CygX3 relative to
that of star~A, as a function of \Xray\ phase (X-ray minimum occurs at
$\phi_{\rm{}X}\simeq0.96$; see Van der Klis \& Bonnet-Bidaud
\cite{vdklb:89}).  The different nights are indicated by different
line types, and are labelled in the figure.  Notice that the 1992 July
25 curve is not passing through the sixth point, since we consider
that point unreliable (see text).  The other points are accurate to
about 10\%.  (Note that it is tacitly assumed here that star~A does
not vary)\label{fig:icont}}
\end{figure}

\paragraph{1992 July 25--27.} 
The source was very weak, the only detectable features being
\HeII\,$(5-4)$ in emission and the interstellar bands in absorption.
The continuum level of the individual spectra, as determined from the
power-law fits described above, is shown as a function of \Xray\ phase
in Fig.~\ref{fig:icont}.  It appears that the continuum is modulated,
with the minimum occurring close to the time expected from infrared
photometry (i.e., the time of \Xray\ minimum) in the second and third
night, but much later in the first night.  In order to check the
reality of the observed variations, we inspected the raw count rates
of star~A in all our spectra.  We found that these were consistent to
within $\sim\!10$\% with the expected variation due to changes in
airmass for all but one spectrum, namely the sixth taken on 1992 July
25, for which the count rate was exceptionally low.  In this spectrum,
the spatial profile looks somewhat extended, suggesting that the stars
were not centred exactly right on the slit.  Therefore, we regard the
corresponding point in Fig.~\ref{fig:icont} as unreliable.  For the
other points, we estimate from the count rates of star~A that the
uncertainties are $\sim\!10$\%. (Note that we tacitly assume that
star~A does not vary.)

\begin{figure}[tb] 
\centerline{\hbox{\psfig{figure=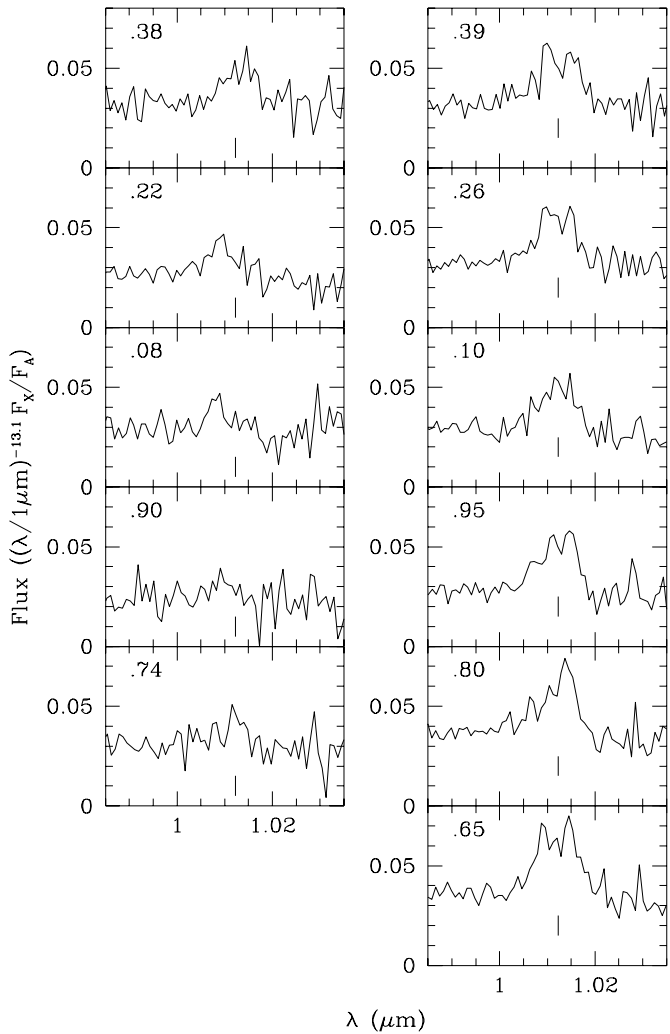,bbllx=32pt,bburx=322pt,width=8.8cm}}}
\caption[]{The I-band spectra around \HeII\,$(5-4)\,\lambda1.0123$
obtained on 1993 June 13 (left) and June 14 (right).
The tickmark near the bottom of each panel indicates the
rest wavelength of \HeII\,$(5-4)$.  The typical error on each point is
$\sim\!0.004$.  The numbers in the top-left indicate the mid-exposure
\Xray\ phase\label{fig:i93}}
\end{figure}

\paragraph{1993 June 13.} 
At this date, the continuum was about 30\% higher than in 1992.
Emission is clearly present at \HeII\,$(5-4)$, and possibly at
\HeII\,$(8-5)$.  The profiles of \HeII\,$(5-4)$ for the individual
spectra are shown in Fig.~\ref{fig:i93} (left-hand panels).  The
centroid wavelength of the profile shows a clear modulation, going
from the rest wavelength at the beginning of the night to a maximum
blueshift of $\sim\!1000\,\kms$ in the third spectrum, and then
shifting back, ending at a redshift of $\sim\!700\,\kms$ in the last
spectrum.  The lines are much less broad than in the following night
and in 1991.  The phasing is similar to what we observed in the K band
in 1992 May, in that maximum blueshift occurs approximately at the
time of \Xray\ minimum (Paper~II; see below).  The continuum level is
also changing during the night (see Fig.~\ref{fig:icont}), showing an
ill-defined minimum near \Xray\ phase~0.

\paragraph{1993 June 14.} 
The spectrum has changed considerably from the previous night.  The
\HeII\,$(5-4)$ and $(8-5)$ emission lines are much stronger, and even
lines corresponding to transitions of \HeII\,$(n-6)$ are present.
Most striking is the appearance of a line at 1.083\,\micron, which we
identify with \HeI\,$2p\,^3{\rm{}P}^0-2s\,^3{\rm{}S}$.  Comparing the
spectrum with spectra of WN stars obtained by Vreux et al.\
(\cite{vreu&a:90}) and by ourselves, we find that the line ratio of
\HeI\,$\lambda1.083$ and \HeII\,$\lambda1.0123$ indicates a subclass
of WN6 or WN7.  In WN stars, the \HeI\ line often shows a P-Cygni
profile, with a strong emission component and a weak absorption
component (see, e.g., Vreux et al.\ \cite{vreu&a:89}).  Unfortunately,
the signal-to-noise ratio of our spectrum is such that we cannot
determine whether a similar absorption feature was present in \CygX3.

In Fig.~\ref{fig:i93} (right-hand panels), the individual profiles of
\HeII\,$(5-4)$ are shown.  No systematic wavelength shifts are
apparent, although the profile does seem to vary.  The continuum level
is slightly higher than in the previous night, and is modulated,
showing a minimum around $\phi_{\rm{}X}=0$ (see Fig.~\ref{fig:icont}).

\subsection{The K-band spectra}

\begin{figure}[tp]
\centerline{\hbox{\psfig{figure=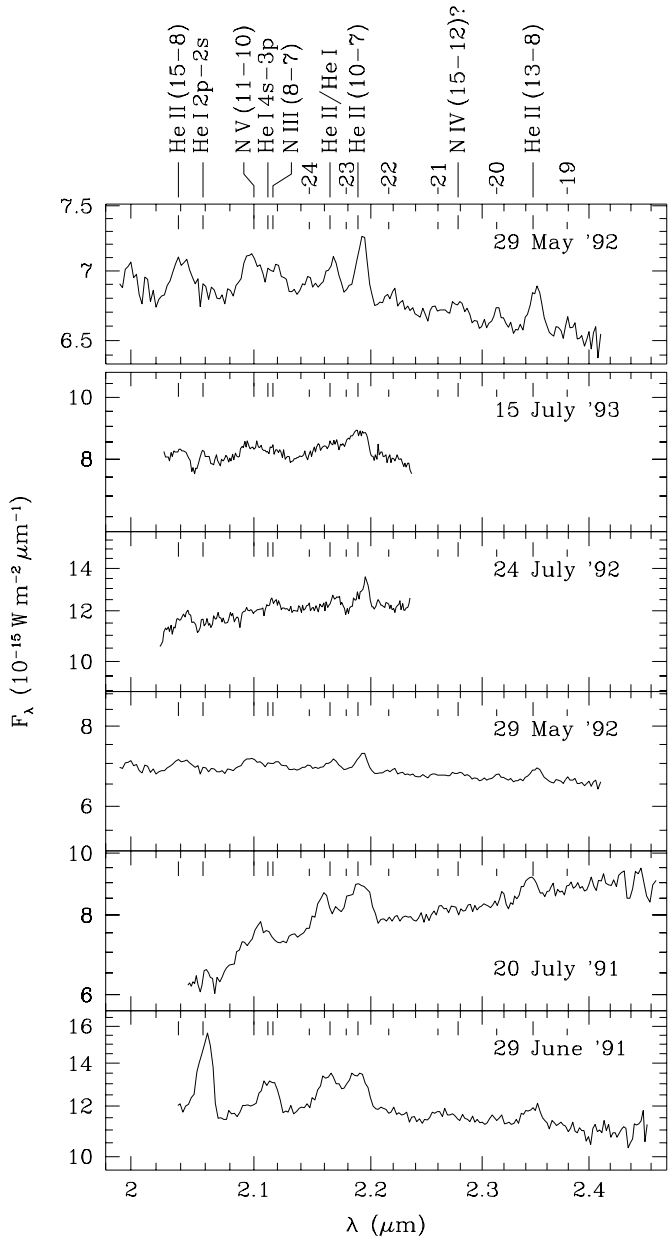,bbllx=32pt,bburx=322pt,width=8.8cm}}}
\caption[]{The K-band spectra of \CygX3.  When more than one spectrum
was taken within one night, the average spectrum is shown.  In the
five lower panels, the spectra are all shown on the same logarithmic
scale, to ease comparison of the continuum slopes and relative line
strengths.  In the top panel, the average of the 1992 May spectra is
shown enlarged.  Line identifications for the stronger lines are given
at the top of the figure.  The line labelled \HeII/\HeI\ is a blend of
\HeII\,$(14-8)$ and members of the \HeI\,$(7-4)$ transition array.
Also indicated are some weaker lines that are present as well in many
of the spectra.  For these, the numbers given refer to the upper
levels of transitions of \HeII\,$(n-9)$\label{fig:kav}}
\end{figure}

The averages of K-band spectra taken in one night are shown in
Fig.~\ref{fig:kav}.  Identifications of the lines are indicated in
the figure and listed in Table~\ref{tab:kid}.

\paragraph{1991 June 29.} 
In this spectrum, first presented in Paper~I, a number of \HeI\ and
\HeII\ emission lines are present, as well as the \NV\,$(11-10)$ line
at 2.100\,\micron\ (not identified in Paper~I).  The spectral type as
indicated by the relative strengths of \HeI\,$\lambda2.112$ and
\HeII\,$(10-7)$ is WN7.  The \HeI\ line at 2.058\,\micron, however, is
anomalously strong for a WN7 star.  This may be due to an absence of
absorption rather than an excess of emission, since WN stars show --
if the line is present -- a P-Cygni profile with a strong absorption
component (Hillier \cite{hill:85}, Williams \& Eenens
\cite{wille:89}).

\paragraph{1991 July 20.} 
The emission-line spectrum is quite different from the first spectrum.
The \HeI\ line at 2.058\,\micron\ has disappeared and the other \HeI\
lines have weakened, while the equivalent widths of the \HeII\ lines
are similar.  On close inspection, it seems that while the \HeII\
lines appear close to their rest wavelengths, the \HeI\ lines are
blue-shifted by about 700\,\kms.

In this spectrum, the continuum is much redder than in all the other
K-band spectra.  We believe this difference is genuine, despite the
fact that the spectrum was taken under mediocre weather conditions
(which may affect the level of the continuum, but is expected to
affect the shape of the spectrum only near the strongest telluric
features, i.e., shortward of 2.1\,\micron).  It is possible that the
redness of the continuum is related to the huge radio outburst that
occurred a few days later, peaking at July 26 (at 8.3\,GHz; Waltman et
al.\ \cite{walt&a:94}).  Dr.~Coe (private communication) provided J,
H, K and narrow-band L photometry taken at UKIRT on 1991 August 6, 14
and 16, when the K-band magnitude was 11.16, 11.91 and 11.85
($\pm0.03$), $J-K$ was 3.48, 3.31 and 3.42 ($\pm0.05$), $H-K$ 1.39,
1.29, and 1.25 ($\pm0.05$), and $K-L$ 1.53, 1.38 and 1.06 ($\pm0.10$),
respectively.  These data appear to show a decline of an outburst,
during which the source is getting less red and fainter.

\begin{figure}[tb]
\centerline{\hbox{\psfig{figure=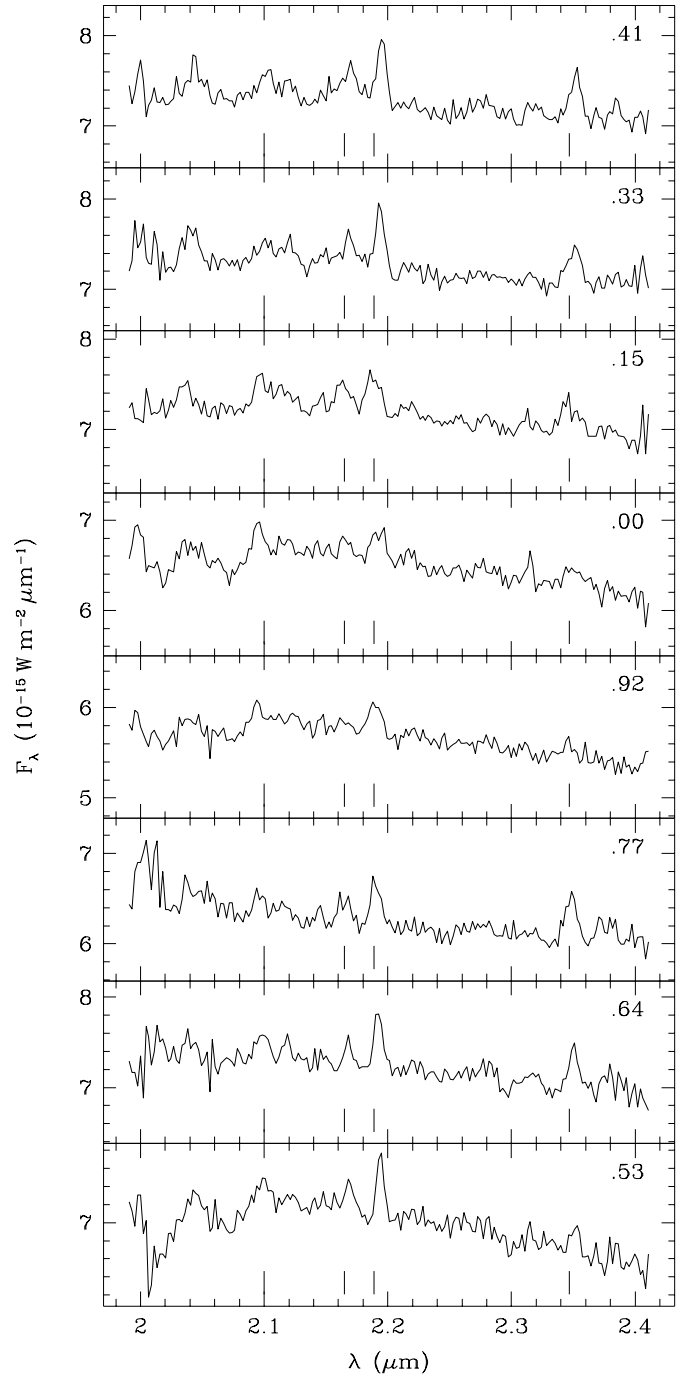,bbllx=32pt,bburx=322pt,width=8.8cm}}}
\caption[]{The K-band spectra obtained on 1992 May 29.  The tickmarks
indicate the rest wavelengths of $\NV\,(11-10)\,\lambda2.100$,
$\HeII\,(14-8)\,\lambda2.165$, $\HeII\,(10-7)\,\lambda2.189$ and
$\HeII\,(13-8)\,\lambda2.347$.  The numbers in the upper right-hand
corner indicate the mid-exposure \Xray\ phase\label{fig:k92}}
\end{figure}

\paragraph{1992 May 29.} 
The series of spectra obtained in this night were presented in
Paper~II.  Compared with the earlier spectra, the lines are much
weaker.  The \HeI\ lines have disappeared, while the \NV\ line at
2.100\,\micron\ and the \NIII\ line at 2.116\,\micron\ are relatively
strong.  One feature in the spectra that we could not readily identify
is the emission line at 2.278\,\micron\ (see Fig.~\ref{fig:kav}).  In
the spectra of WN stars that we have available, there is also an
emission feature at this wavelength, which we attribute to
\NIV\,$(15-12)$, but in all spectra it is weaker than the neighbouring
\HeII\,$(21-9)$ line.  However, compared to what is seen in WN stars,
the \NV\ line is also exceptionally strong relative to the \HeII\
lines.  Therefore, we tentatively identify the 2.278\,\micron\
feature with \NIV\,$(15-12)$.

In Fig.~\ref{fig:k92}, the individual spectra are shown.  As discussed
in Paper~II, there are clear wavelength shifts during the night, with
maximum blueshift coinciding with infrared minimum, and maximum
redshift half an orbit later.  The modulation of the continuum is
similar to what has been found previously from photometric studies.
The good observing conditions allowed us to determine a light curve
(Paper~II, Fig.~3; see also Fig.~\ref{fig:ffstudy}) from the
individual pairs of integrations that form the spectra (see
Sect.~\ref{sec:obsk}).

\paragraph{1992 July 24.} In this spectrum, the lines are somewhat 
stronger than in May, and the degree of ionisation is somewhat lower,
although not as low as in 1991.  The lines are all shifted by about
800\,\kms\ to the red.  The profile of \HeII\,$(10-7)$ is rather
asymmetric, with a blue wing.  The night was not photometric, and the
uncertainty in the flux level is about 20\%.

\begin{figure}[tb]
\centerline{\hbox{\psfig{figure=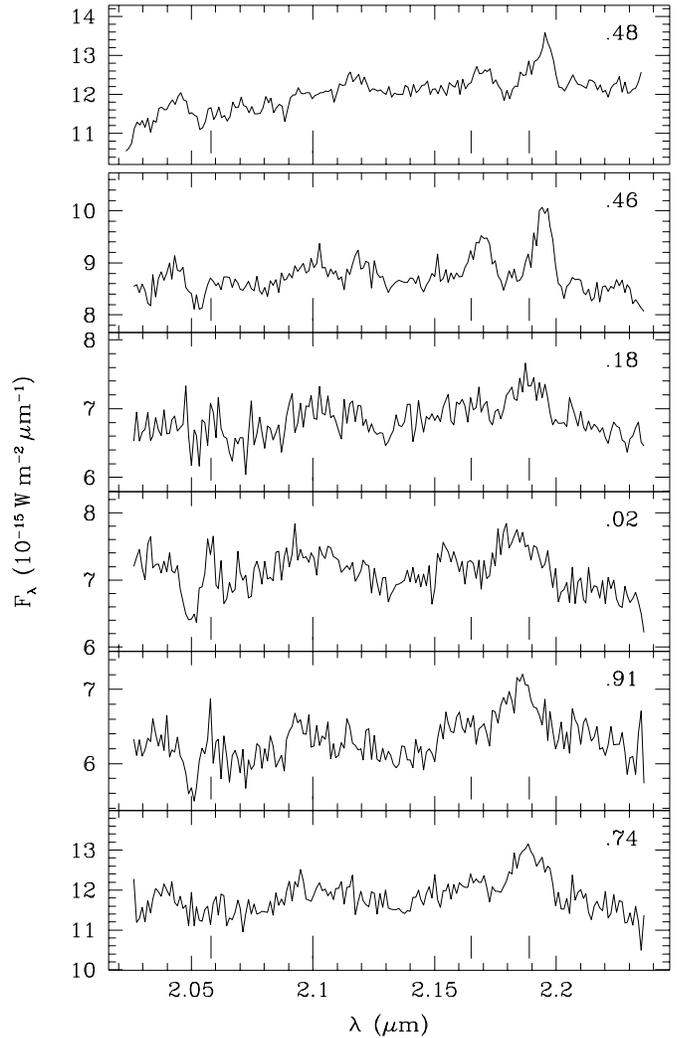,bbllx=32pt,bburx=322pt,width=8.8cm}}}
\caption[]{The K-band spectra obtained on 1993 July 15.  For
comparison, the spectrum taken on 1992 July 24, is shown in the top
panel.  The tickmarks indicate the rest wavelengths of
$\HeI\,\lambda2.058$, $\NV\,(11-10)\,\lambda2.100$,
$\HeII\,(14-8)\,\lambda2.165$ and $\HeII\,(10-7)\,\lambda2.189$, and
the numbers in the upper right-hand corner the mid-exposure \Xray\
phase.  Notice that the scale has been chosen such that the relative
scale is the same for each panel, contrary to Fig.~\ref{fig:k92}.
This is because due to the mediocre observing conditions, the absolute
fluxes are not reliable for these spectra\label{fig:k93}}
\end{figure}

\paragraph{1993 July 15.} 
A striking difference between the spectra obtained in this night and
the other K-band spectra is the appearance of \HeI\,$\lambda2.058$ in
absorption.  In the individual spectra (Fig.~\ref{fig:k93}), an
absorption feature at $\sim\!2.05\,\micron$ is clearly seen in the
second and third spectrum (close to \Xray\ phase 0) and possibly in
the fifth.  

The line strengths shown in the spectra taken this night, are similar
to those observed in 1992 July.  The lines are modulated, phased in a
similar fashion as in 1992 May.  The line profiles show marked
asymmetries, as is most clearly seen in the \HeII\,$(10-7)$ line.  The
shapes of the lines in the fifth spectrum are similar to those
observed in 1992 July, at about the same \Xray\ phase (see
Fig.~\ref{fig:k93}), even though the equivalent widths of the lines
are somewhat smaller at the latter date.  In addition to a modulation
in wavelength, the shape of the lines seems to be modulated, the line
width being smallest at the time of maximum redshift.  From the
fluxes, it seems that the continuum level is modulated as well,
reaching minimum near $\phi_{\rm{}X}=0$.  Notice, however, that due to
the mediocre observing conditions the uncertainty on the flux levels
is about 20\%, Therefore, it is not possible to construct a light
curve as could be done for the 1992 May data.

\begin{table*}
\caption[ ]{I-band line identifications and equivalent widths$^{\rm{}a}$
\label{tab:iid}}
\begin{tabular}{lllllllll@{}}
\hline
$\lambda_{\rm lab}$& Identification& 1991 June 21& 1992 June 25& 
1992 June 26& 1992 June 27& 1993 June 13& 1993 June 14& Average\\
(\micron)& & (\AA)& (\AA)& (\AA)& (\AA)& (\AA)& (\AA)& (\AA)\\ 
\hline
\multicolumn{8}{l}{\em Emission}\\
0.9345& \HeII\,$8-5$& 
    \pho9(5)& \water & \water & \notdet    &\pho9(6)&\pho15(5)\\
1.0123& \HeII\,$5-4$&
      90(20)& 35(20) & $<8$   & 11(5)      &  29(10)&\pho76(8)\\
1.0830& \HeI\,2$p^3$P$^0-2s^3$S& 
      \nodata & \notdet& \notdet& \notdet    &    ?   &  140(30)\\
$\ldots$& \HeII\,$n-6^{\rm b}$&
    \notdet & \notdet& \notdet& \notdet    & \notdet& $n\leq19$\\[1mm]
\multicolumn{8}{l}{\em Absorption}\\
\doublebox!0.8620!0.8649!& 
  i.s.\ bands&    11(3)&\pho9(2)&\pho9(3)&\pho8(3)&11(4)& \phoo6(3)&9(2)\\
\hline
\end{tabular}
\vskip1mm\noindent
$^{\rm a}$ The listed equivalent widths are for the average spectra of
the six nights, plus the equivalent width of the interstellar bands
for the average of all spectra.  The minus signs for the equivalent
widths of emission lines have been omitted.  Errors (values in
brackets) and upper limits indicate 90\% confidence levels.  Further
indications are: ?: possibly present; \water: spectrum shows apparent
feature due to an overcorrection of the telluric \water\ absorption
features; \notdet: not detected; \nodata: not in spectral range\\
$^{\rm b}$ Lines with upper levels up to the indicated number are
present\\
\end{table*}

\begin{table*}
\caption[ ]{K-band line identifications and equivalent widths$^{\rm a}$
\label{tab:kid}}
\begin{tabular}{lllllllllllll@{}}
\hline
$\lambda_{\rm lab}$& Identification& Range$^{\rm a}$&
\multicolumn{2}{l}{1991 June 29}& \multicolumn{2}{l}{1991 July 20}&
\multicolumn{2}{l}{1992 May 29}& \multicolumn{2}{l}{1992 July 24}&
\multicolumn{2}{l}{1993 July 15}\\
(\micron)&&(\micron)& (\AA)&&(\AA)&&(\AA)&&(\AA)&&(\AA)&\\
\hline
\multicolumn{13}{l}{\em Emission}\\
2.038& \HeII\,$15-8$&           &  \nodata &&                
                                   \nodata &&                
                                   \pho4(2) C&&              
                                   \pho5(2)&&                
                                   \pho5(2)&\\[1mm]          
2.058& \HeI\,2$p\,^1$P$^0-2s\,^1$S&&
                                      45(5)&&                
                                   \notdet &&                
                                   \notdet &&                
                                   \notdet &&                
                                   \pho2(1)&\\[1mm]          
2.100& \NV\,$11-10$&           &           & 1/5&            
                                           & 1/3&            
                                           & 3/5&            
                                           & 2/5&            
                                           & 3/5           \\
\doublebox!2.112!2.113!& 
\doublebox!\HeI\,4$s\,^3$S$-3p\,^3$P$^0$!\HeI\,4$s\,^1$S$-3p\,^1$P$^0$!&
                     2.07--2.13&      22(3)&\doublebox! !4/5 B!\pho& 
                                      15(5)&\doublebox! !2/3 B!\pho& 
                                      11(2)&&                
                                      11(3)&&                
                                      21(3)&\doublebox! !2/5 B!\pho\\
2.116& \NIII\,$8-7$&           &           &&                
                                           &&                
                                           & 2/5 N\pho&          
                                           & 3/5 N\pho&          
                                           &\\[1mm]          
2.165& \HeI\,$7-4$&             &          & 1/5&            
                                           & 1/5&            
                                           &&                
                                           &&                
                                           &               \\
2.165& \HeII\,$14-8$& 2.15--2.20&     61(5)& 1/5&            
                                      58(5)& 1/5&            
                                      15(3)& 2/5&            
                                      19(3)& 1/3&            
                                      43(5)& 1/3           \\
2.189& \HeII\,$10-7$&           &          & 3/5&            
                                           & 3/5&            
                                           & 3/5&            
                                           & 2/3&            
                                           & 2/3\\[1mm]      
2.279& \NIV\,$15-12$&           &     $<4$ &&                
                                   \notdet &&
                                   3.5(15) &&                
                                   \nodata &&
                                   \nodata &\\[1mm]
2.347& \HeII\,$13-8$&           &    12(3) &&                
                                     12(3) &&                
                                      7(2) &&                
                                   \nodata &&
                                   \nodata &\\[1mm]
$\ldots$&\HeII\,$n-9^{\rm b}$&  &\multicolumn{2}{l}{$n\leq24$}&
                                 \multicolumn{2}{l}{$n\leq22$}&
                                 \multicolumn{2}{l}{$n\leq24$}&
                                 \multicolumn{2}{l}{$n\leq24$}&
                                 \multicolumn{2}{l}{$n\leq24$}\\[1mm]
\multicolumn{13}{l}{\em Absorption}\\
2.058& \HeI\,2$p\,^1$P$^0-2s\,^1$S&  
                                &  \notdet && 
                                   \notdet && 
                                   \notdet &&
                                         ? &&
                                    1.6(6) &\\[1mm]
\hline
\end{tabular}
\vskip1mm\noindent 
$^{\rm a}$ The listed equivalent widths are determined from the
average spectrum for each of the five nights.  The minus signs for the
equivalent widths of emission lines have been omitted.  In case of
line blends, the equivalent widths were determined for the whole
blend, using the wavelength range indicated in the column labelled
`Range'.  For each date, the left-hand column lists the equivalent
width, and the right-hand column the estimated fractional
contributions for lines that are part of a blend.  Errors (values in
brackets) and upper limits indicate 90\% confidence levels.  Further
indications are: ?: possibly present; C: the continuum surrounding the
line is depressed due to residual telluric \cotwo\ absorption; B:
combined contribution of \HeI\ and \NIII; N: most likely, the main
contribution is from \NIII; \notdet: not detected; \nodata: not on
spectrum\\ 
$^{\rm b}$ lines with upper levels up to the indicated number are
present\\
\end{table*}

\subsection{Characteristics of the variations\label{sec:varchar}}

From Figs.~\ref{fig:iav} and \ref{fig:kav}, it is clear that the
spectral appearance of the source is highly variable on time scales
longer than a few orbital periods.  Still, in all spectra, most of the
emission features can be readily identified with features expected for
Wolf-Rayet stars of the nitrogen-rich WN subclass, i.e., with lines of
different ions of helium and nitrogen (see
Figs.~\ref{fig:iav},~\ref{fig:kav}; Tables~\ref{tab:iid} and
\ref{tab:kid}).  We compared the spectra with those of normal
Wolf-Rayet stars (Van Kerkwijk \cite{vker:95}; see also Hillier et
al.\ \cite{hill&a:83}; Hillier \cite{hill:85}; Vreux et al.\
\cite{vreu&a:89}, \cite{vreu&a:90}), as well as with the spectra of
two early-type Of stars that show WN-like infrared spectra (Conti et
al.\ \cite{cont&a:95}).  We find that the emission at 2.167\,\micron\
is much weaker relative to the other lines than in the Of stars and
those WN stars that contain significant amounts of hydrogen.

When the lines are strong, the spectra are most similar to those of
stars of spectral type WN6/7, although the lines are somewhat weaker
than average.  The line equivalent widths are similar to those of
WR\,78 (WN7), WR\,153 (\hbox{WN6-A}; here, the suffix A indicates a
subclass of WN stars that show weak -- equivalent width smaller than
40\AA\ -- emission at \HeII\,$\lambda5411$) and WR\,155 (WN7).  Note
that both WR\,153 and WR\,155 are binaries (see Vreux et al.\
\cite{vreu&a:90}), so that their line strengths may be smaller due to
the continuum contribution of the companion.  One line that is
different is \HeI\,$\lambda2.058$.  In WN stars, this line, if
present, usually has an absorption component that is stronger than the
emission component (e.g., Hillier \cite{hill:85}; Williams \& Eenens
\cite{wille:89}), while in \CygX3 it was observed strongly in emission
on 1991 June 29, and in absorption near \Xray\ phase 0 on 1993 July
15.

When the lines are weak, their relative strengths indicate an earlier
subclass of WN4/5.  The line-strength ratios are inconsistent with
this spectral type, however, in that the nitrogen lines are too strong
relative to the \HeII\ lines.  In both the I and the K-band spectra,
the equivalent widths of the \HeII\ lines are about a factor ten
smaller than is found for normal WN stars.  The \NV\,$(11-10)$ line at
2.100\,\micron\ is about a factor 3 to 4 weaker than in WR\,5 (WN5;
Hillier et al.\ \cite{hill&a:83}) and WR\,128 (WN4; Van Kerkwijk
\cite{vker:95}).  The line at 2.279\,\micron, which we have
tentatively identified with \NIV\,$(15-12)$, is about as strong as it
is in WR\,128.  In the spectrum of WR\,5, the feature does not seem to
be present.

When the lines are weak, they shift in wavelength as a function of
orbital phase, with maximum blueshift occurring at \Xray\ and infrared
minimum (note that \Xray\ minimum occurs at $\phi_{\rm{}X}\simeq0.96$;
see Van der Klis \& Bonnet-Bidaud \cite{vdklb:89}), and maximum
redshift half an orbit later.  In the three runs in which we observed
this, we find a maximum redshift of about 700\,\kms, while the maximum
blueshift is between 300\,\kms\ in the 1992 May K-band spectra and
1000\,\kms\ in the 1993 June I-band spectra.  Note that the redshift is
easier to measure than the blueshift, as the shape of the lines is
varying as well, the profile at maximum redshift being narrower,
stronger and more asymmetric than that at other phases (we will
discuss this in terms of our model in Sect.~\ref{sec:modelresults}).

From the spectra, we find that the continuum level between different
observations varies by up to a factor~2, while the continuum slope is
rather constant (except on 1991 July 20; see above).  The average
continuum level appears to be correlated with the line strength, the
level being lower when the lines are weaker.

The variations of the continuum level as a function of phase that we
observed in the K-band on 1992 May 29, are consistent with earlier
photometric results (Becklin et al.\ \cite{beck&a:73},
\cite{beck&a:74}; Mason et al.\ \cite{maso&a:76}, \cite{maso&a:86};
see Paper~II).  For the other series of K-band spectra, taken on 1993
July 15, we can only state that the time of the minimum is consistent
with the expected phase.  In the I band, the light curves we find are
more irregular than those in the K band, but mostly of similar
appearance, in contrast to the large variety of behaviour observed by
Wagner et al.\ (\cite{wagn&a:89}) from I-band photometry.

\section{An X-ray ionised Wolf-Rayet wind\label{sec:model}}

\begin{figure}[tb]
\centerline{\hbox{\psfig{figure=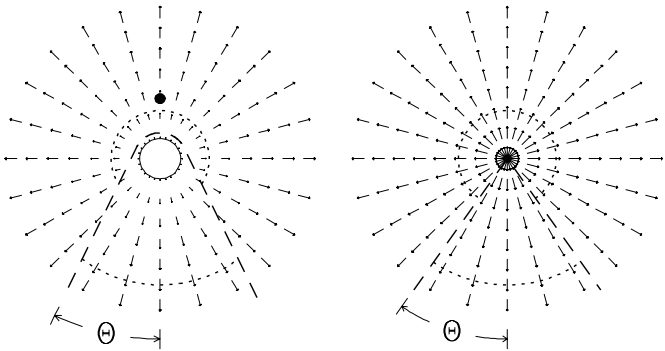,angle=90,width=8.8cm}}}
\caption[]{The model of \CygX3 (taken from Paper~II).  On the left,
a schematic representation of the system is drawn. The compact
object is indicated by a black dot and the helium star by an open
circle. The arrows indicate the (accelerating) wind of the helium
star. The region not ionized by the \Xray{}s originating from the
compact object is bounded by the long-dashed curve.  The opening angle
with respect to the \Xray\ source is $\Theta$.  The short-dashed
curves indicate the characteristic radii of the 2.2\micron\ emitting
region.  On the right, the simplified model used for the
numerical calculations is shown. The assumptions are that the helium
star and its wind can be represented as a constant-velocity wind
originating in the centre of the helium star, and that the shadowed
part of the wind can be assumed to be within a cone with opening angle
$\Theta$ and with the centre of the helium star at its
vertex\label{fig:model}}
\end{figure}

The wavelength shifts shown in the spectra taken on 1992 May 29, were
interpreted in Paper~II in terms of the helium-star model of Van den
Heuvel \& De Loore (\cite{vdhedl:73}; see Sect.~\ref{sec:intro}).  It
was proposed that both the weakness of the lines and the wavelength
shifts resulted from the fact that the Wolf-Rayet wind of the helium
star was highly ionised by the X~rays from the compact object, except
for the part in the \Xray\ shadow of the helium star, and that the
line emission originated mostly from the latter, shadowed part of the
wind (see Fig.~\ref{fig:model}, taken from Paper~II).  In this way,
one naturally obtains the observed phasing of the wavelength shifts,
since at \Xray\ minimum, which occurs at superior conjunction of the
\Xray\ source, the wind in the shadowed part will be moving towards
us, and hence we will observe a blueshift.  Similarly, at \Xray\
maximum, we will observe a redshift.

The spectra presented in this paper confirm the proposed model in two
ways: (i) in both the I and K bands, we observe similar wavelength
shifts in additional series of spectra that show weak lines; and (ii)
for the one series of spectra that show strong lines, we find that
such wavelength shifts are not present.

It was argued in Paper~II that the observed orbital modulation of the
infrared flux of \CygX3, as well as the virtual independence of the
average infrared flux on the degree of ionisation, was a result of
free-free absorption dominating the infrared opacity.  In the
Rayleigh-Jeans tail of the spectrum, the free-free absorption
coefficient, corrected for stimulated emission, is proportional to
$T^{-3/2}$.  Since, in the Rayleigh-Jeans tail, the source function is
proportional to $T$, the total flux will be proportional to $T^{-1/2}$
for an optically thin cloud.  For a wind that is optically thick at
the base, as is the case in the infrared for Wolf-Rayet winds (e.g.,
Hillier et al.\ \cite{hill&a:83}), this anticorrelation with
temperature of the flux emitted per unit volume is compensated by the
positive correlation with temperature of the fraction of the wind
contributing to the observed flux (since at higher temperature more of
the wind is optically thin).  In fact, from an analytical derivation,
Wright \& Barlow (\cite{wrigb:75}; see also Davidsen \& Ostriker
\cite{davio:74}) found that for an isothermal wind that is expanding
with a constant velocity, the free-free flux is to first order
independent of the temperature.

For \CygX3, one expects that, given this lack of sensitivity to the
temperature, the average infrared flux will not depend strongly on
whether or not a large fraction of the wind is hot due to the \Xray\
ionisation: the heated part will be less opaque, but it will emit
about as much infrared emission as it would were it cool, since its
smaller effective emitting area is compensated by its higher
temperature (we will come back to this in
Sect.~\ref{sec:ramifications}).  However, the more opaque, cool part
of the wind in the shadow of the helium star can (partly) obscure the
smaller but brighter hot part.  This will lead to a modulation of the
infrared flux as a function of orbital phase, with a minimum occurring
when the cool part is in front, i.e., at the time of maximum blueshift
of the lines and of \Xray\ minimum.

In order to test the ideas presented above quantitatively, one should
calculate the effects of the presence of the \Xray\ source on the
Wolf-Rayet wind in a self-consistent way.  This would be an iterative
process.  For instance, the ions that are responsible for the
acceleration of the wind could be brought to a higher degree of
ionisation by the X~rays so that the acceleration stops, which would
cause an enhanced wind density that, in turn, might influence the
\Xray\ luminosity.  Numerical calculations of this sort have been made
for \Xray\ binaries such as Vela~X-1, in which the companion is an OB
supergiant (e.g., Blondin et al.\ \cite{blon&a:90}, \cite{blon&a:91}).
From these calculations, it follows that gravity, rotation, radiation
pressure and \Xray\ heating all influence the flow of the wind past
the compact object, leading for instance to unsteady accretion wakes.
At the present time, similar calculations for an \Xray\ source in a
Wolf-Rayet wind do not seem feasible, if only because it is not yet
known how to calculate the wind of a {\em single} Wolf-Rayet star in a
self-consistent way.  Therefore, we have chosen instead to construct a
model in which only the essence of the idea is kept, viz., that the
continuum and lines are formed in a dense wind which has a
two-temperature structure.

\subsection{Description of the model\label{sec:modeldescription}}

The assumptions that we have made for our model are (see
Fig.~\ref{fig:model}): (i) the cool part of the Wolf-Rayet wind is
confined within a cone with a given opening angle and with the helium
star at the vertex; (ii) the temperature of the wind is constant
within both the hot and the cool parts of the wind; and (iii) the wind
originates at the centre of the helium star, and expands at a constant
velocity that is the same in the hot and the cool part of the wind.
In essence, assumptions (i) and (iii) are equivalent to the assumption
that the size of the system and the size of the region where the wind
is accelerated are negligible compared to the size of the region where
the bulk of the flux originates.  A qualitative discussion of the
effects of the breakdown of these assumptions is given in
Sect.~\ref{sec:ramifications} below.

For the calculation of the continuum flux, we assume that the dominant
opacity source is free-free absorption, i.e., we neglect the
contribution of bound-free absorption, as well as the effects of
electron scattering (we will return to this in
Sect.~\ref{sec:ramifications}).  For the calculation of the line
profile, we use the Sobolev approximation (Sobolev \cite{sobo:60};
Castor \cite{cast:70}).  Furthermore, we assume that the line is in
local thermodynamical equilibrium (LTE) throughout the wind.  For
lines that arise from high levels of excitation -- such as the \HeII\
and \NV\ lines -- this is likely to be the case, since, under the
prevailing conditions, the population of these levels relative to the
population in the next ionisation stage will be close to the
population expected in LTE (Griem \cite{grie:63}; Hillier et al.\
\cite{hill&a:83}).  We also assume that the fraction of the ions that
is in the next ionisation stage, is the same in the hot and the cold
part of the wind.  For the \HeII\ lines, this assumption is
reasonable, since helium is already almost fully ionised in normal
Wolf-Rayet winds, and thus -- assuming that the same holds for the
cool part of the wind -- the fraction of helium that is in the form of
\HeIII\ cannot be much higher in the hot part.  However, for the \NV\
lines this assumption may break down.  (Note that the assumption of
LTE is almost certainly invalid for the \HeI\ lines, since these arise
from low levels of excitation.)

With the above-mentioned assumptions, most of the integrations
necessary to calculate the continuum flux and the line profile can be
done analytically.  The derivations are similar to those given by
Wright \& Barlow (\cite{wrigb:75}; continuum) and Hillier et al.\
(\cite{hill&a:83}; line profile) for the case of an isothermal,
constant-velocity, spherically-symmetric wind (we will refer to such a
wind as a `one-temperature' wind).  Details about the derivations are
given by Van Kerkwijk (\cite{vker:93b}).  A point to note is that the
free-free opacity scales as $T^{-3/2}$, while the line opacity is
proportional to $T^{-5/2}$.  Hence, the equivalent width of a line is
anticorrelated with the temperature (see Hillier et al.\
\cite{hill&a:83}).

\subsection{Comparison with normal Wolf-Rayet stars\label{sec:wrcomp}}

For a one-temperature wind, a continuum energy distribution of the
form $F_\nu\propto\nu^{2/3}$ -- where $F_\nu$ is the flux in frequency
units -- is predicted for the Rayleigh-Jeans part of the spectrum
(Wright \& Barlow \cite{wrigb:75}).  This result has been confirmed
observationally both in the radio and infrared bands (Abbot et al.
\cite{abbo&a:86} and references therein).  From a recent study of the
continuum energy distributions between 0.1 and 1\,\micron\ of 78
Wolf-Rayet stars by Morris et al.\ (\cite{morr&a:93}), it appears that
also at smaller wavelengths, the continuum energy distribution is
similar.  Morris et al.\ (\cite{morr&a:93}) found that longward of
0.15\,\micron\ the continua could be well described by a power law of
the form $F_\nu\propto\nu^\beta$.  They argue that this power law can
be extended further into the infrared for all stars that do not show
contributions from dust or nebulae.  They found that the frequency
distribution of the values of $\beta$ was approximately Gaussian, with
a mean value of $\beta$ of 0.85 and a standard deviation of 0.4 (which
they attribute to intrinsic star-to-star differences).  Thus, the
assumptions we make in our model seem to be reasonable for the
infrared continuum.

\begin{figure}[tb]
\centerline{\hbox{\psfig{figure=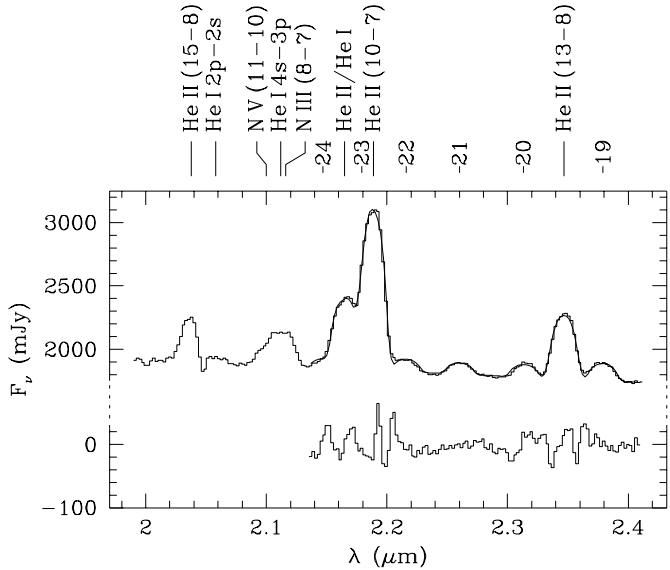,bbllx=32pt,bburx=322pt,width=8.8cm}}}
\caption[]{The K-band spectrum of WR\,134 (WN6).  Superposed is the
best-fitting spectrum for an isothermal, constant-velocity,
spherically symmetric wind.  The lower curve shows the difference on a
magnified scale.  The spectrum has not been fitted shortward of
2.14\,\micron, since for the \HeI\ and \NIII\ lines present there, the
assumption of LTE is most likely invalid (see text).  Note that the
strength of each line was a free parameter, i.e., not determined from
the condition of LTE
\label{fig:wr134fit}}
\end{figure}

To verify the applicability of the assumptions underlying the
line-profile calculations, we fitted our K-band spectrum of WR\,134 --
a Wolf-Rayet star of spectral type WN6 that has broad, well-resolved
lines -- using a power-law for the continuum, and line profiles as
expected for a one-temperature wind (Hillier et al.\
\cite{hill&a:83}).  The fitted parameters are the level and power-law
index of the continuum, the wind velocity, the velocity shift due to
the radial velocity of the star (as well as to possible errors in the
wavelength calibration), and, for all lines, values for a scaling
parameter proportional to the line opacity that determines the line
strength (no attempt has been made to determine the values of these
scaling parameters from the condition of LTE; the main purpose here is
to demonstrate that the shape of the observed line profiles can be
reproduced with the one-temperature model).  Furthermore, the
interstellar reddening to the source ($E_{B-V}=0.47\pm0.02$; Morris et
al.\ \cite{morr&a:93}), the seeing, the slit width and the detector
pixel size are taken into account.  The result is shown in
Fig.~\ref{fig:wr134fit}.  For the slope of the continuum, we found a
value of $\beta\simeq0.8$, consistent with the value of $0.76\pm0.10$
found by Morris et al.\ (\cite{morr&a:93}) for $\lambda<1\,\micron$.
The velocity of the wind that we find is $\sim\!1700\,\kms$, similar
to the 1900\,\kms\ listed by Schmutz et al.\ (\cite{schm&a:89}) and
Prinja et al.\ (\cite{prin&a:90}).  As can be seen in
Fig.~\ref{fig:wr134fit}, the line profiles are satisfactorily
reproduced.  We conclude that the assumption that the \HeII\
lines are in LTE is adequate for calculating the line profiles arising
in the two-temperature wind of \CygX3.

\begin{figure*}[tb]
\centerline{\hbox{\psfig{figure=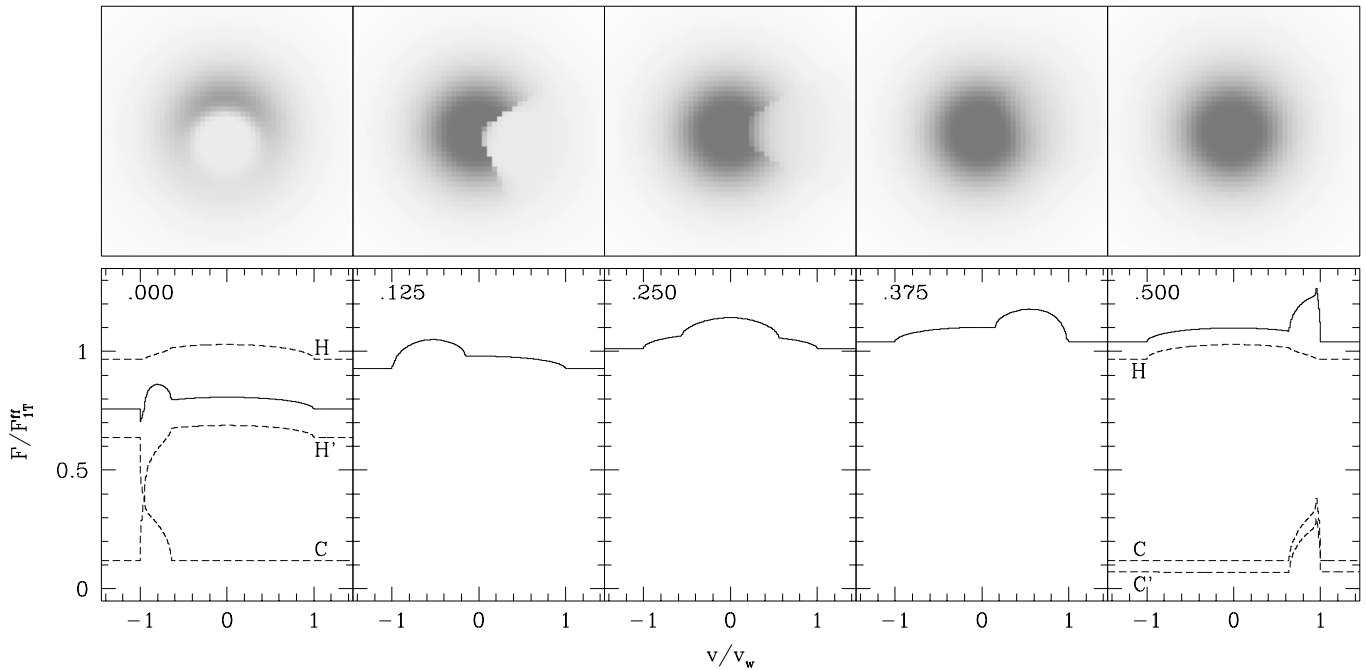,width=18cm}}}
\caption[]{The orbital-phase dependent variations expected for a
two-temperature wind.  The upper row of panels shows the predicted
surface brightness due to the free-free (continuum) emission which one
would observe were one able to resolve the source.  The grey-scale
transformation used is linear between no flux (white) and about twice
the maximum flux (black).  The cone of less bright material that one
sees in the centre of the three left-hand panels is the cool part of
the wind.  In the left-most panel, it obscures the bright centre
region of the hot part of the wind.  For the calculation, the opening
angle of the cone was~$34^\circ$, the inclination~$74^\circ$, and the
temperature ratio~7.  The lower row of panels shows the predicted
integrated free-free and line flux (full curve), relative to the
free-free flux expected for a one-temperature wind.  The weak, broad
component of the line profile originates in the hot part of the wind,
and the strong, sharp one in the cool part.  If the whole wind were
cool, the line peak would be at $\sim\!40$\% above the continuum.  In
the leftmost and rightmost panels, the dashed curves labelled $H$ and
$C$ show the emission from, respectively, the hot and the cool part of
the wind, as one would observe it if the other part of the wind were
not present.  The curve labelled $H'$ shows the emission from the hot
part as observed through the cool part, and $C'$ the cool part as
observed through the hot part\label{fig:phases}}
\end{figure*}

\subsection{Model results and comparison with the 
observations\label{sec:modelresults}} 

For the calculation of the free-free flux arising in a two-temperature
wind, the following parameters are needed: (i)
$F^{\rm{}ff}_{\rm{}1T}$, the flux that a one-temperature wind would
produce; (ii) $i$, the inclination of the system; (iii) $\Theta$, the
opening angle of the cone within which the wind is cool; and (iv)
$Q_T$, the temperature ratio between the hot and the cold parts of the
wind.  For the line profiles, the additional parameters are (v)
$v_{\rm{}w}$, the velocity of the wind; and (vi) $K^{\rm{}line}$, a
scaling parameter proportional to the line opacity.  Furthermore, for
the comparison with the observations, one also needs (vii) $t_0$, the
time of superior conjunction of the \Xray\ source.

As an example, we show in Fig.~\ref{fig:phases} (bottom panels) the
modelled line plus continuum flux, relative to the continuum flux
expected for a one-temperature wind, for a number of orbital phases
(measured with respect to the time of superior conjunction of the
\Xray\ source), with the parameters taken from Paper~II ($Q_T=7$,
$i=74^\circ$ and $\Theta=34^\circ$).  Also shown (top panels) is the
modelled continuum surface brightness that one would observe if one
were able to resolve the source.  In the latter, one can see that, in
the centre, the hot part of the wind (the one which extends over a
larger angle) is brighter (blacker) than the cool part.  This is
because in these dense regions of the wind, the optical depth is large
both in the hot and in the cool part, so that the surface brightness
is determined by the Planck function (i.e., proportional to the
temperature in the Rayleigh-Jeans tail).  Far away from the centre,
the surface brightness in the cool part is slightly larger (this is
just noticeable in, e.g., the right-hand side of the third panel),
since at these distances the wind is optically thin, and the lower
temperature is more than compensated for by the larger optical depth.
In the leftmost panel, one can see the blocking of the brightest
region of the hot part of the wind by the much more opaque cool part.

\begin{figure*}[tb]
\parbox[b]{11.2cm}{\psfig{figure=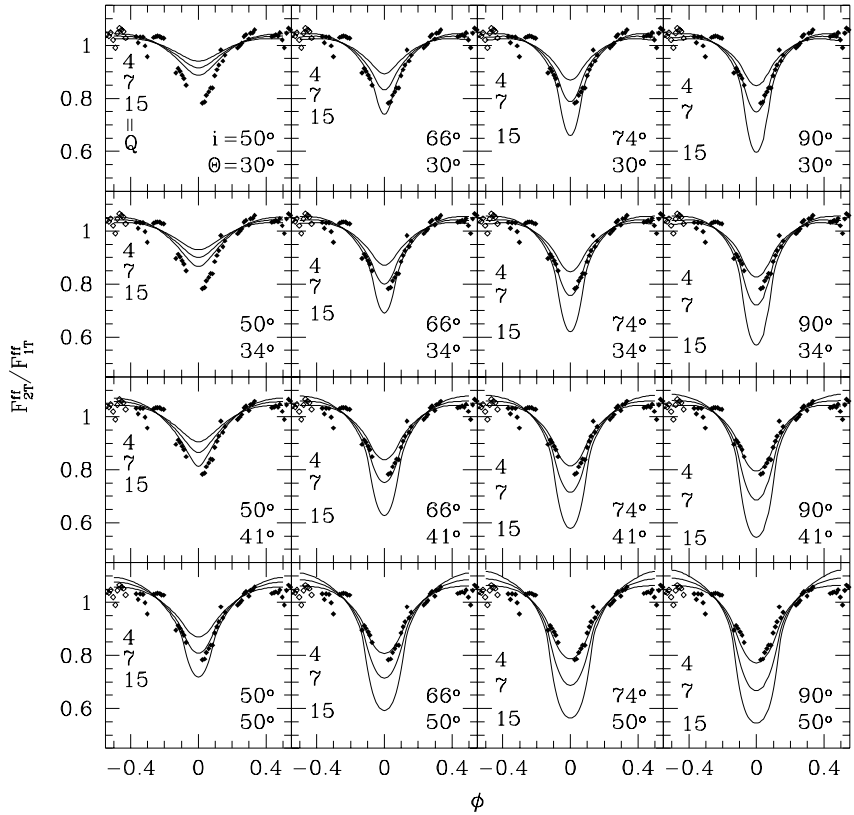,bbllx=32pt,bburx=400pt,width=11.2cm}}\hfill
\parbox[b]{6cm}{%
\caption[]{The model infrared light curve.  In the panels, light
curves are drawn as a function of orbital phase for $Q_T=4$, 7 and 15
for different values of $i$ and $\Theta$ (as indicated).  The observed
light curve -- obtained from the 1992 May K-band spectra (see
Paper~II) -- is overdrawn (filled symbols).  Some points are shown
shifted by one orbital period as well (open symbols).  For all points,
the observing times were converted to orbital phase using a time of
superior conjunction $t_0={\rm{}JD_{bar}}\,2448771.997$, and the
fluxes were divided by $F^{\rm{}ff}_{\rm{}1T}=11.373\,$mJy.  These
values of $t_0$ and $F^{\rm{}ff}_{\rm{}1T}$ are the best-fit values
for $Q_T=7$, $i=74^\circ$, and $\Theta=34^\circ$ (for the other sets
of parameters, they are not very different).  The time $t_0$
corresponds to \Xray\ phase $\phi_{\rm{}X}=0.87$.  For reference,
\Xray\ minimum occurs at $\phi_{\rm{}X}=0.96$ (Van der Klis \&
Bonnet-Bidaud \cite{vdklb:89})\label{fig:ffstudy}}}
\end{figure*}

The line profile is composed of two components that shift in velocity
with orbital phase.  The broad weak component arises in the hot part
of the wind, and the sharp strong one in the cool part.  The latter is
rather weak at orbital phase~0, when the \Xray\ source is at superior
conjunction and the cool part of the wind is moving most directly
towards us (see Fig.~\ref{fig:model}).  The line profile even shows a
small absorption feature.  This is because in the line the opacity in
the cool part of the wind is larger, and hence the absorption of the
continuum emission from the hot part of the wind is more effective.
Conversely, half an orbit later, when the cool part is moving away
from us, the line component arising in it is rather strong.  In fact,
the red wing is stronger than would be the case if the whole wind were
cool, because the intervening continuum opacity in the hot part of the
wind is lower than would be the case for a completely cool wind.  

In Fig.~\ref{fig:ffstudy}, the predicted infrared light curves are
shown for a range of values of the parameters $(Q_T,i,\Theta)$.  For
comparison, the light curve derived from the 1992 May spectra is
overdrawn (see Paper~II), using $F^{\rm{}ff}_{\rm{}1T}=11.373\,$mJy
and $t_0={\rm{}JD_{bar}}\,2448771.997$ as derived from fitting the
lightcurve to the model with $Q_T$ held fixed at~7 (Paper~II).  The
best-fit values of these two parameters hardly vary with the choice of
$(Q_T,i,\Theta)$.  From the figure, it is clear that the infrared
light curve does not uniquely determine $Q_T$, $i$ and~$\Theta$.  The
parameter combinations that yield the most satisfactory fit to the
observed light curve are those with with
$(Q_T,i,\Theta)=(4,90^\circ,41^\circ)$, $(7,74^\circ,34^\circ)$ and 
$(15,66^\circ,30^\circ)$.
\begin{figure*}[tb]
\parbox[b]{11.2cm}{\psfig{figure=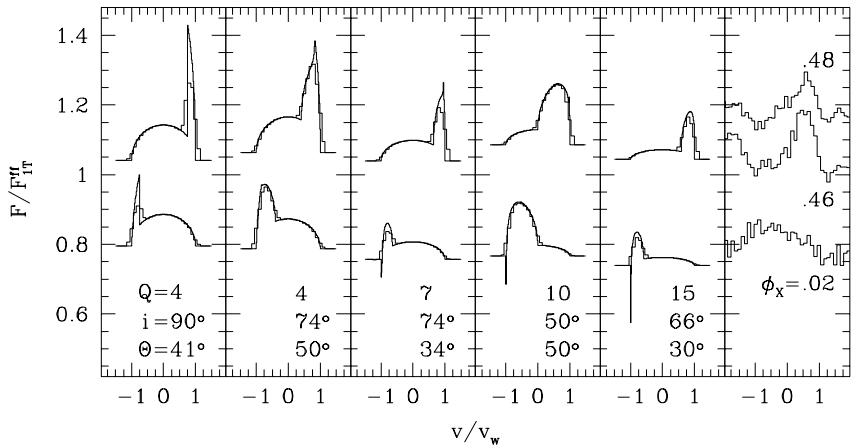,bbllx=32pt,bburx=400pt,width=11.2cm}}\hfill
\parbox[b]{6cm}{%
\caption[]{Observed and modelled Line profiles at maximum blueshift
(lower curves) and at maximum redshift (upper curves).  In the
rightmost panel, three observed \HeII\,$(10-7)$ profiles are shown
(taken from Fig.~\ref{fig:k93}).  The fluxes of the three profiles
have been divided by 10.5, 8.5 and
$9\,10^{-15}\,{\rm{}W\,m^{-2}\,\micron^{-1}}$ (top to bottom), and for
the conversion to fractional velocity, a wind velocity of 1500\,\kms\
was used.  In the other panels, model profiles are shown for five
different sets of parameters (as indicated).  The drawn curve reflects
the model results, and the histogram the results binned to the
resolution of the observations (no smoothing over phase was
applied)\label{fig:diffffline}}}
\end{figure*}

In Paper~II it was shown that the wavelength shifts observed in 1992
May could be reproduced with the model.  The wavelength shifts shown
in the other observations are very similar to these, and therefore we
will not discuss these further, but rather focus on the line shapes
predicted by the model, and compare these with the profiles observed
in the higher resolution K-band spectra that we now have available.

In Fig.~\ref{fig:diffffline}, we show the observed profiles for
maximum redshift and maximum blueshift, and compare these with the
predicted profiles.  The latter were calculated for the three
parameter sets that produce the best fits to the observed light curve
(see above), as well as for two other sets that produce adequate fits
and that represent somewhat different positions in parameter space.
From the figure, one notices that the modelled lines are weaker at
maximum blueshift than at maximum redshift, as discussed above for the
example shown in Fig.~\ref{fig:phases}.  Small absorption components
at maximum blueshift are found in three cases.  Their presence
reflects the geometry of the model: they are present for values of
$\Theta$ slightly larger than $(90^\circ-i)$, i.e., for the case in
which a significant fraction of the cool matter moving almost directly
towards us, is seen projected against the hot part of the wind.  For
$\Theta$ significantly larger than $(90^\circ-i)$, the optically thick
central region of the cool part of the wind intervenes.

In comparing the modelled profiles with the observed ones in more
detail, one should keep in mind that $K^{\rm{}line}$, which determines
the strength of the line, has been kept the same for all models.  For
the value chosen, the line would peak at 40\% above the continuum if
the whole wind were cool.  The equivalent width would be
$0.6\,v_{\rm{}w}$, or $\sim\!70(v_{\rm{}w}/1500\,\kms)\,\AA$ at
\HeII\,$(10-7)$ (where 1500\,\kms\ is the velocity indicated by both
the \HeI\ and \HeII\ line profiles; see Sect.~\ref{sec:massloss}
below).  In WN6/7 stars, equivalent widths ranging from 30 to
130\,\AA\ are observed, with the stronger lines occurring in WN6 stars
(Hillier et al.\ \cite{hill&a:83}; Hillier \cite{hill:85}).  Hence,
differences in the line strength of the order of a factor~2 are to be
expected.  Keeping this in mind, some conclusions can be drawn from
the comparison with the observed profiles.  Specifically, the relative
strength shown by the model profiles at maximum redshift and the
relative weakness at maximum blueshift, is seen in the observed \HeII\
profiles as well, both in the series of K-band spectra obtained in
1992 (Fig.~\ref{fig:k92}) and the one obtained in 1993
(Fig.~\ref{fig:k93}).  Furthermore, the asymmetry of the observed
profiles is consistent with that expected from the two-temperature
structure of the wind.

From a closer comparison (Fig.~\ref{fig:diffffline}), one sees that
the observed profiles -- which are clearly resolved -- are best
reproduced for relatively large values of the opening angle and for
values of the inclination not too close to $90^\circ$ (note, however,
that, given the integration times of 20--30 minutes, the observed
profiles represent averages over a range of about 0.1 in orbital
phase, and thus are expected to be somewhat broader than the modelled
ones).  From the different profiles somewhat different conclusions
about the temperature ratio $Q_T$ are inferred.  The lower of the two
maximum-redshift profiles in Fig.~\ref{fig:diffffline} shows hardly
any blue-shifted emission, and from this profile one would thus infer
a high value of $Q_T$.  In contrast, the maximum-blueshift profile
(taken in the same run) and the other maximum-redshift profile (taken
a year earlier) show profiles from which a lower value $Q_T$ would be
indicated.  Given both the difficulty in determining the observed
ratio due to the uncertainty in the continuum level, and the
simplicity of the model, it seems that differences like these are to
be expected, and that, in general, the characteristic variations shown
in the observed profiles (Sect.~\ref{sec:varchar}) can be adequately
reproduced.

\section{Wind velocity and mass-loss rate\label{sec:massloss}}

For a constant-velocity, isothermal wind, the free-free flux depends
only on the ratio of the mass-loss rate $\dot M$ and the velocity
$v_{\rm{}w}$ of the wind ($\dot M/v_{\rm{}w}\propto F^{3/4}$; Wright
\& Barlow \cite{wrigb:75}).  In Paper~II, the relations given by
Wright \& Barlow (\cite{wrigb:75}) and Hillier et al.\
(\cite{hill&a:83}) were used to estimate the mass-loss rate in \CygX3.
With the parameter $F^{\rm{}ff}_{\rm{}1T}$ determined from the fit to
the 1992 May infrared light curve, corrected for 1.5 magnitudes of
interstellar extinction as derived by Becklin et al.\
(\cite{beck&a:72}), and the wind velocity determined from the
wavelength shifts, a mass-loss rate $\dot{M}_{\rm{}ff}=
4\,10^{-5}\,\Msun\,{\rm{}yr^{-1}}$ was derived (for a distance of
10\,kpc, and assuming completely ionised helium and a gaunt factor of
unity; cf.~Hillier et al.\ \cite{hill&a:83}).

Another estimate of the mass-loss rate can be made on the basis of the
observed increase of the orbital period, as was done in Paper~I, if
one assumes that the specific angular momentum carried away by the
wind is equal to that of the helium star.  With the orbital parameters
of Kitamoto et al.\ (\cite{kita&a:95}), we find $\dot{M}_{\rm{}dyn}=
M_{\rm{}tot}\dot{P}/2P=
0.6\,10^{-5}(M_{\rm{}tot}/10\,\Msun)\,\Msun\,{\rm{}yr^{-1}}$, where
$M_{\rm{}tot}$ is the total mass of the system.  Hence, for reasonable
values of the mass of the helium star ($\sim\!10\,\Msun$) and the
compact object ($1.4\,\Msun$), $\dot M_{\rm{}dyn}$ is smaller than
$\dot M_{\rm{}ff}$ by about a factor of~6.  This discrepancy may well
be larger, since, as shown below, the present observations indicate
that $\dot M_{\rm{}ff}$ may have been underestimated in Paper~II.

In Paper~II, a wind velocity of 1000\,\kms\ was used to estimate $\dot
M_{\rm{}ff}$.  From the observations presented here, it is possible to
derive a more accurate estimate of the wind velocity.  Williams \&
Eenens (\cite{wille:89}) have argued that a reliable estimate of the
terminal velocity of the wind can be obtained from the blue edge of
blue-shifted absorption features of the \HeI\,$\lambda2.058$ line,
such as we have observed in the K-band spectra taken on 1993 July 15
(Fig.~\ref{fig:k93}).  For both spectra in which the feature is
unambiguously detected (the second and third), we find a centroid
velocity of $-1200\pm100\,\kms$.  The full widths at half maximum are
$800\pm150$ and $1000\pm200\,\kms$, respectively.  Hence, for the blue
edge of the absorption feature we find, after correction for the
resolution of the spectra ($\sim\!450\,\kms$), $-1500\pm200$ and
$-1600\pm250\,\kms$ for the second and third spectrum, respectively.
(Here, possible effects of turbulence have been neglected.  See
Williams and Eenens (\cite{wille:89}).)

Another estimate of the wind velocity can be obtained from the widths
of the lines at times that they do not show an orbital modulation.  In
Paper~I, it was found that the \HeII\ lines in the 1991 I and K-band
spectra had full widths at half maximum (FWHM) consistent with a
single value of 2700\,\kms, with an estimated uncertainty of
200\,\kms.  For the \HeII\,$(5-4)$ lines in the spectra taken on 
1993 June 14, we find a FWHM of $2800\pm300\,\kms$.  In order
to find the velocity to which these widths correspond, we have to make
an assumption about the intrinsic line profile.  Torres et al.\
(\cite{torr&a:86}) find that, for a line with a parabolic shape, the
conversion factor $C_{\rm{}vel}\equiv v_{\rm{}w}/$FWHM equals 0.71,
while for a line with a rectangular profile, it is 0.5.  For the
profile of a line in LTE arising in a one-temperature wind, we find
that $C_{\rm{}vel}=0.57$.  Of course, these estimates only apply for
well-resolved lines.  For our spectra, with a resolution of
$\sim\!500\,\kms$ in the I band, and $\sim\!900\,\kms$ in the K band,
we find that $C_{\rm{}vel}$ for the one-temperature profile should be
increased to $\sim\!0.6$.  Using $C_{\rm{}vel}=0.6\pm0.1$ as a
conservative estimate, we find that the observed FWHMs of the \HeII\
lines correspond to a wind velocity $v_{\rm{}w}=1650\pm300\,\kms$.

The maximum velocity shown by the modulated profiles can also be used
to estimate the wind velocity.  In the 1993 K-band spectra,
(Fig.~\ref{fig:k93}; also Fig.~\ref{fig:diffffline}) emission is seen
out to a redshift of $1500\pm150\,\kms$.  After correction for the
resolution of $\sim\!450\,\kms$, we find a velocity of
$1450\pm150\,\kms$.

The estimates mentioned above indicate a wind velocity of
$\sim\!1500\,\kms$, rather than the 1000\,\kms\ used in Paper~II.
This velocity is consistent with that expected for a WN6/7 Wolf-Rayet
star (e.g., Eenens \& Williams \cite{eenew:94}).  If the wind velocity
is the same in the continuum-forming regions (i.e., if the assumption
of a constant velocity is valid), then with the new velocity the
mass-loss rate mentioned above will increase by a factor~1.5.

Another uncertainty in the estimate of the mass-loss rate stems from
the variability by up to a factor~2 of the phase-averaged infrared
intensity of the source.  From the spectra that we have available now,
it seems likely that the intensity of the source is systematically
lower when the lines are weak.  In terms of our model, the wind is
most `Wolf-Rayet like' when the lines are strong, and therefore the
mass-loss rate might be underestimated if one uses, as in Paper~II,
the flux estimate derived from the weak-line 1992 May spectra.  Adopting
the flux observed in the 1991 June spectrum, the
inferred mass-loss rate would increase by another factor~1.5.

For the interstellar extinction $A_{\rm{}K}$ in the K band, a value of
1.5 magnitudes was used.  This value was derived by Becklin et al.\
(\cite{beck&a:72}) from the infrared colours of \CygX3, under the
assumption that the intrinsic colour $H-K\la0.3$.  However, as we will
show below (Sect.~\ref{sec:fcont}), the shape of the infrared
continuum indicates $A_{\rm K}\simeq2.0$.  Hence, the intrinsic flux
may have been underestimated in Paper~II.  For an increase of 0.5
magnitudes in $A_{\rm{}K}$, the estimate of the mass-loss rate
increases by a factor~1.4.

In summary, the estimate of the mass-loss rate $\dot{M}_{\rm{}ff}$
given in Paper~II should be considered as a lower limit.  If all three
corrections mentioned above were to apply, $\dot M_{\rm{}ff}$ would
increase by a factor~3, to $1.2\,10^{-4}\,\Msun\,{\rm{}yr}^{-1}$, an
order of magnitude larger than the dynamical estimate
$\dot{M}_{\rm{}dyn}$.

Interestingly, for WR\,139 (V444~Cyg), a Wolf-Rayet binary for which a
dynamical estimate of the mass-loss rate is available, this estimate
is lower by about a factor~3 than the estimates based on the free-free
radio emission and on the modelling of the infrared spectral lines
(St-Louis et al.\ \cite{stlo&a:93}, and references therein).  For this
system, St-Louis et al.\ (\cite{stlo&a:93}) find additional evidence
for the lower mass-loss rate from polarisation measurements.  They
suggest that the discrepancy might result from inhomogeneities in the
wind.  Since the free-free and (non-resonance) line emission processes
depend on the square of the density, these inhomogeneities would lead
to enhanced free-free and line fluxes, and hence, when the
inhomogeneities are neglected, the mass-loss rate will be
overestimated.  In contrast, the estimates based on the polarisation
depend linearly on the density, and thus are independent of these
inhomogeneities (provided the wind is optically thin to electron
scattering).

Additional evidence that the mass-loss rates in Wolf-Rayet stars may
be lower than is inferred from line and free-free continuum fluxes,
comes from the profiles of strong \HeII\ lines (e.g.,
\HeII\,$(3-2)\,\lambda1640$, \HeII\,$(7-4)\,\lambda5411$).  For these
lines, the electron-scattering wings predicted by model calculations
tend to be stronger than those observed by a factor of $\sim\!2$
(Hillier \cite{hill:90}).  Hillier (\cite{hill:90}) suggests that this
discrepancy, like the one mentioned above, reflects the presence of
inhomogeneities in the wind, due to which the line emission processes
-- which scale with the density squared -- are enhanced, while the
electron scattering process -- which scales linearly with the density
-- is not.  In order to test this expectation, Hillier
(\cite{hill:91}) performed exploratory model calculations on an
inhomogeneous wind.  He found that, indeed, lines as strong as those
obtained for a homogeneous wind, but with electron-scattering wings in
accordance with the observations, could be obtained, for a mass-loss
rate that was about half that needed for a homogeneous wind.

From the above, we conclude that the mass-loss rate $\dot M_{\rm{}ff}$
of $\sim\!10^{-4}\,\Msun\,{\rm{}yr}^{-1}$ that is inferred from the
infrared flux, should be regarded as a highly uncertain estimate of
the genuine mass-loss rate $\dot M_{\rm{}true}$.  From the cases
discussed above, it follows that $\dot M_{\rm{}true}$ may be smaller
than $\dot M_{\rm{}ff}$ by a factor of 2--3.  For \CygX3, it seems not
unlikely that the difference is even larger, since possible
inhomogeneities may well be enhanced as a result of its variable
nature.  The conclusions drawn in the previous section are not
necessarily affected, however, provided the inhomogeneity of the wind
is similar in both the hot and the cold part of the wind.

\section{Ramifications\label{sec:ramifications}}

In this section, we discuss qualitatively the complications that arise
for cases in which one or more of the assumptions underlying our model
are invalid.  We start with the main assumption, viz., that the size
of the continuum and line emitting regions is much larger than both
the orbital separation and the size of the regions of the wind where
significant velocity and temperature gradients are present
(Sect.~\ref{sec:modeldescription}).  Thereafter, we briefly discuss
the effects of the rotation of the system, of the contribution of
bound-free processes to the absorption opacity, and of electron
scattering.

\subsection{The size of the continuum and line emitting
regions\label{sec:sizes}}

Wright \& Barlow (\cite{wrigb:75}) define the characteristic radius of
the free-free emitting region by the condition that the integrated
flux emitted exterior to that radius (i.e., without optical-depth
effects) equals the flux from the whole wind (including optical-depth
effects).  At this radius, the radial optical depth to infinity is
0.244.  It is given by
\begin{eqnarray}
R^{\rm ff}_{\rm char}&=8.8&
  \left(\frac{\dot M_{\rm ff}}{10^{-4}\,\Msun\,{\rm yr}^{-1}}\right)^{2/3}
  \left(\frac{v_{\rm w}}{1500\,\kms}\right)^{-2/3}\times\nonumber\\ 
&&\left(\frac{\lambda}{2.2\,\micron}\right)^{2/3}
  \left(\frac{T}{5\,10^4\,{\rm K}}\right)^{-1/2} 
  \,\Rsun.\label{eq:rff}
\end{eqnarray} 
Here, the mass-loss rate $\dot{M}_{\rm{}ff}$ and the wind velocity
$v_{\rm{}w}$ were scaled to the values found in
Sect.~\ref{sec:massloss}, and the temperature $T$ to an estimate of
the temperature in the cool part of the wind.  We have used a
mass-loss estimate of $10^{-4}\,\Msun\,{\rm{}yr^{-1}}$ for
$\dot{M}_{\rm{}ff}$, which we argued above might be influenced by the
presence of inhomogeneities, since, for the estimate of the free-free
radius, we want to include these effects.  Note that the effective
radius of the wind, defined by $F^{\rm ff}=4\pi B(T)R_{\rm{}eff}^2$,
is at a radial optical depth to infinity of 0.05 (Hillier et al.\
\cite{hill&a:83}), and that it is a factor 1.7 larger than the
characteristic radius defined above.

From the temperature dependence of the characteristic radius, it is
clear that the assumption that it is large compared to the other
relevant sizes, will break down first in the high-temperature part of
the wind.  Taking the size of the Roche lobe of the helium star --
1.7\,\Rsun\ for a 10\,\Msun\ helium star and a 1.4\,\Msun\ compact
object (Paper~I) -- as an optimistic estimate of the value of the
characteristic radius below which the assumptions will break down, we
find that the critical value $Q_{\rm{}crit}$ of the temperature ratio
is~27 in the K band (for the values of $\dot{M}_{\rm{}ff}$,
$v_{\rm{}w}$ and $T$ used in Eq.~\ref{eq:rff}).  As a more pessimistic
estimate, the orbital separation can be used, which, for the masses
quoted above, is 3.2\,\Rsun.  The value of $Q_{\rm{}crit}$
corresponding to this radius is~8.  For the I band ($0.9\,\micron$),
the values for the two cases are $Q_{\rm{}crit}=8$ and 2.3,
respectively.

The estimates of the critical values of $Q_T$ are within the range of
values that we have used for our model.  Hence, it is likely that
deviations from the model results discussed above occur.  For the wind
of the helium star, one expects that both the assumption that the wind
is isothermal, and the assumption that is has a constant velocity,
will not be valid close to the star.  Furthermore, the assumption that
the shadow can be represented as a cone with the helium star at its
vertex, will break down.  We will discuss these three points in turn.

\paragraph{Temperature.} 
Close to the surface of the helium star, where the X~rays cannot
penetrate, the temperature of the wind will be determined by the
helium-star properties alone, and thus be lower than assumed in our
model.  To obtain a crude estimate of the consequences, we consider
the case where outside of a certain radius $R_{\rm{}b}$ the wind is
strongly heated, and within that radius not at all.  At very long
wavelengths, the heated part of the wind will be optically thick, and
our model assumptions apply.  However, going to shorter wavelengths,
the heated part will become optically thin, and the cooler layers
below will become visible.  In these layers, the optical depth is much
higher, and thus these layers will effectively appear as a blackbody
of radius $R_{\rm{}b}$, with the temperature determined by the helium
star properties.  Hence, the spectrum will flatten, then steepen at
shorter wavelengths as the wind ceases to contribute, and finally
flatten at still shorter wavelengths when the cooler layers become
transparent.

\paragraph{Velocity.} 
For high values of $Q_T$, one will observe the deeper, accelerating
layers of the wind (see Fig.~\ref{fig:model}).  Since the velocity in
these layers is lower, the density and thus the free-free opacity will
be higher.  In consequence, the effects will be opposite to those
discussed for the temperature above: the flux will be higher, and,
since this effect will be stronger at shorter wavelengths, the
spectrum will be steeper than is expected for a constant-velocity
wind.

\paragraph{Geometry} 
The assumption that the wind is confined within a cone with the helium
star at its vertex, will also cease to be realistic for high values
of~$Q_T$.  From trial calculations on a two-temperature wind with a
genuine shadow (see Fig.~\ref{fig:model}), we found that, indeed, for
high values of $Q_T$, the light curves differ significantly (mostly
showing a wider minimum), due to the very dense parts of the wind,
well within a stellar radius of the star, now becoming shadowed and
thus cool.  However, since close to the star the assumption of a
constant temperature and a constant velocity will also be invalid (see
above), we do not consider these results as better than the ones
obtained with our model in its simplest geometric form.

\subsection{Rotation\label{sec:rotation}}

Due to the orbital motion in the system, the matter in the wind of the
helium star will start to trail the star at larger radii.  Hence, in
assuming that the cool part is within a cone, we implicitly assume
that both the cooling in the shadow, and the \Xray\ heating outside
are very fast.  If this is not the case, the cool part of the wind
will be lagging at larger radii.  An observational indication that
this may in fact happen, is that the \Xray\ light curve shows an
asymmetric minimum, with a rather steep ingress and a more
moderately-sloped egress.  In the infrared, a similar effect may be
present.  For instance, the infrared light curve observed by Mason et
al.\ (\cite{maso&a:76}), shows an ingress that is steeper than the
egress.  The K-band light curve we observed, is more symmetric
(Fig.~\ref{fig:ffstudy}), but from the points in the phase interval
$-0.26$ to $-0.2$, it does appear that ingress is somewhat faster than
egress.

\subsection{Bound-free opacity\label{sec:boundfree}}

Another effect that we have neglected in our model, is the
contribution of the bound-free process to the opacity.  Since this
contribution decreases with increasing temperature, one may expect
that, going from a one-temperature to a two-temperature state, the
infrared flux will decrease.  Hillier et al.\ (\cite{hill&a:83}) found
that, for a temperature of 50000\,K, the contribution of the
bound-free opacity to the total opacity is $\sim\!15\%$ of that of the
free-free opacity in the K band and $\sim\!45\%$ in the I band.  At
30000\,K -- a reasonable lower limit to the temperature in the cool
part -- the relative contributions are $\sim\!25\%$ and $\sim\!80\%$,
respectively.  However, Hillier et al.\ also note that the effect is
almost cancelled by the error made in assuming the Rayleigh-Jeans
approximation.  (For instance, at 0.9\,\micron\ and 50000\,K, both the
black-body flux and the free-free opacity are about 15\% lower than
expected in the Rayleigh-Jeans approximation, leading to a 25\%
decrease in expected flux.)  Hence, it seems unlikely that with the
inclusion of bound-free opacity in the model, one would be able to
reproduce the observed changes in the continuum of up to a factor 2 in
both the I and K bands.

The contribution of the bound-free opacity might also reflect itself
in an increased depth of the modulation, since, due to its
contribution, the cool part of the wind should be more opaque, while
the hot part should not.  Given the wavelength dependence of this
contribution, any effect should be stronger at shorter wavelengths,
and one would, therefore, expect that the depth of the modulation
would increase going to shorter wavelengths.  The observational
evidence on this is confused: Fender \& Bell-Burnell (\cite{fendb:94})
find that the depth of the modulation is larger at H than at~K, but
Molnar (\cite{moln:88}) finds that there are no colour changes with
orbital phase.

\subsection{Electron scattering\label{sec:scattering}}

Electron scattering will be important if the radius at which
$\tau_{\rm{}es}\simeq1$ exceeds the characteristic radius
$R^{\rm{}ff}_{\rm{}char}$ of the free-free emitting region.  For
constant-velocity, spherically-symmetric mass loss, this radius is
given by
\begin{equation}
R_{\rm es}=9.6
  \left(\frac{\dot M}{10^{-4}\,\Msun\,{\rm yr}^{-1}}\right)
  \left(\frac{v_{\rm w}}{1500\,\kms}\right)^{-1}\,\Rsun.
\label{eq:res}
\end{equation}
Hence, for the free-free mass-loss rate of
$\sim\!10^{-4}\,\Msun\,{\rm{}yr}^{-1}$ (Sect.~\ref{sec:massloss}),
electron scattering will be important in both the hot and the cold
parts of the wind.  However, as discussed in Sect.~\ref{sec:massloss},
the genuine mass-loss rate $\dot M_{\rm{}true}$, with which the
electron scattering optical depth scales, is most likely lower by at
least a factor 2--3.  For such a lower estimate,
$R_{\rm{}es}\simeq3$--5\,\Rsun, and electron scattering will be
important in the hot part of the wind for $Q_T\ga3$--8.  Due to the
electron scattering, the hot part of the wind will appear more
extended.  Hence, the depth of the modulation will decrease somewhat,
and the minimum will become somewhat wider.

At the radial distance of the \Xray\ source ($a\simeq\!3\,\Rsun$; see
above), the electron scattering optical depth to infinity will be
$\ga\!1$ even when the mass-loss rate estimate is decreased by a
factor 2--3.  Thus, the \Xray\ source will effectively be larger, and
the shadow of the helium star will become less well-defined,
especially at larger radii.  One may expect that, in these less
shadowed regions, the degree of ionisation will be increased.  This
may explain the absence of \HeI\ emission in the spectra with weak,
modulated lines.

\section{The continuum energy distribution\label{sec:fcont}}

If the hypothesis that the infrared continuum and lines in \CygX3
originate from a Wolf-Rayet wind, is correct, it is to be expected
that the continuum energy distribution is similar to that observed in
Wolf-Rayet stars, i.e., that it can be described by a power-law of the
form $F_\nu\propto\nu^\beta$ (Morris et al.\ \cite{morr&a:93}; see
Sect.~\ref{sec:modeldescription}).  To test this expectation, and also
to look for possible deviations that might result from the break down
of the assumptions underlying our model (as discussed above), we
collected photometric data from the literature, and combined these
with the our spectroscopic results.  For this purpose, the K-band
spectra can be used directly, since these are flux calibrated.  This
is not the case for our I-band spectra.  However, since these were
reduced relative to the nearby star~A (see Sect.~\ref{sec:obsi}), for
which photometry is available, we can obtain an estimate of the fluxes
using star~A as a local standard.

\begin{figure*}[tb]
\parbox[b]{11.2cm}{\psfig{figure=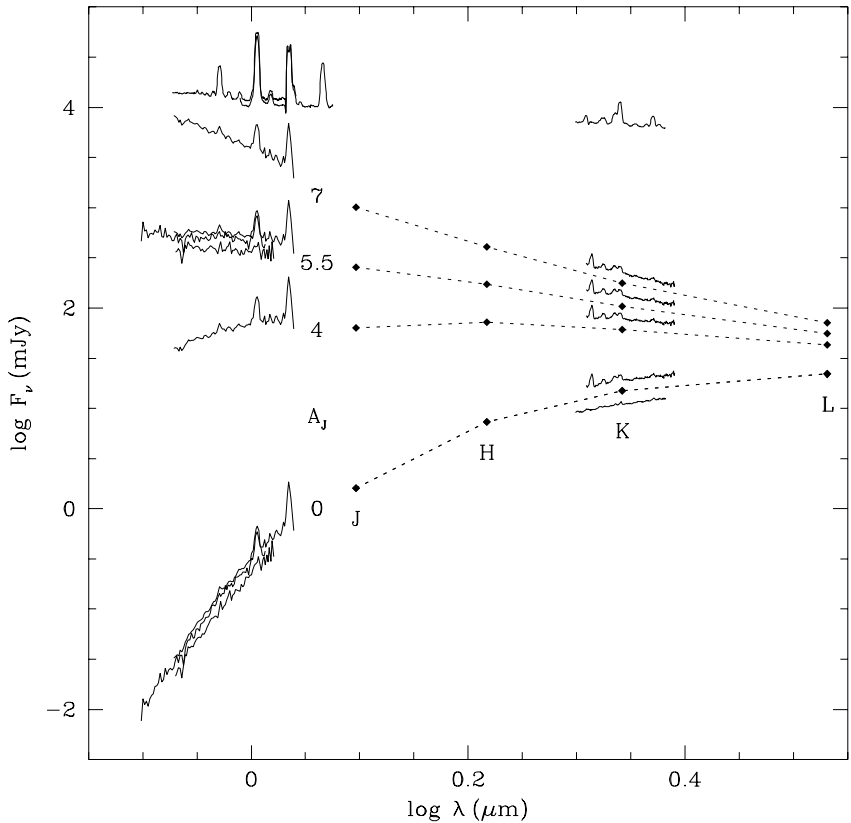,bbllx=32pt,bburx=400pt,width=11.2cm}}\hfill
\parbox[b]{6cm}{%
\caption[]{The continuum flux distribution.  Shown are the fluxes
derived from the JHKL photometry of Molnar (\cite{moln:88}), and from
our spectra.  The lower set of points and curves ($A_{\rm{}J}=0$)
reflects the observed data.  The JHKL points are connected by a dashed
line to indicate that they were obtained simultaneously.  Two K-band
spectra are shown, the spectrum with strong lines obtained on 1991
June 29, and the weak-line average spectrum obtained on 1992 May 29.
In the I-band, the strong-line average spectrum obtained on 1993 June
14 is shown, the weak-line average spectrum obtained on 1992 July 25,
and, to extend the wavelength range to smaller wavelengths, the
strong-line spectrum obtained on 1991 June 21 (notice that, for the
latter, the flux level may be underestimated; see
Sect.~\ref{sec:obsi}).  The other sets of points and curves reflect
the data corrected for the interstellar extinction for three different
values of $A_{\rm{}J}$ (as indicated).  The extinction law we used was
that of Mathis (\cite{math:90}).  For the IJHKL bands, this extinction
law gives $A_\lambda\propto\lambda^{-1.7}$.  For comparison, the
flux-calibrated I, J and K-band spectra of WR\,134 (WN6), corrected
for interstellar extinction using $A_{\rm J}=0.69$ (Morris et al.\
\cite{morr&a:93}), are shown at the top (shifted upward by 0.5\,dex
for clarity)\label{fig:isabs}}}
\end{figure*}

For star~A, Westphal et al.\ (\cite{west&a:72}) list $V=15.05$,
$B-V=0.98$ and $U-B=0.43$.  From our spectra, we found that the
spectral type is in the range F5 to~G0 (Sect.~\ref{sec:obsi}).  From
the observed colours, we conclude that the star cannot be a
supergiant.  For main-sequence stars of the above-mentioned range in
spectral type, the effective temperatures range from 6400 to 5900\,K,
and the intrinsic colours $\{(U-B)_0,(B-V)_0,(V-I)_0\}$ range from
$\{0.00,0.43,0.64\}$ to $\{0.11,0.59,0.81\}$ (Johnson \cite{john:66}).
With the interstellar extinction law given by Mathis (\cite{math:90}),
the $E_{B-V}$ and $E_{U-B}$ colour excesses correspond to values of
the interstellar extinction at~J ranging from $A_{\rm{}J}=0.48$ to
0.34, and from $A_{\rm{}J}=0.52$ to 0.39, respectively.  [Note that
the interstellar extinction law of Mathis (\cite{math:90}) is
normalised at~J.  This band is chosen because longward of
$\sim\!1\,\micron$ the extinction law is virtually identical in all
lines of sight.] Hence, the expected Johnson I magnitude ranges from
13.52 to~13.59.

For the flux calibration, we assume that, in the wavelength range
covered by our spectra, star~A can be represented by a black body of
6000\,K, reddened by interstellar extinction corresponding to
$A_{\rm{}J}=0.4$, and scaled to produce an I-band magnitude of 13.55
(these values are appropriate for spectral type~F8).  To estimate the
accuracy of this procedure, we also calibrated our WHT I-band spectra
of the Wolf-Rayet stars, and compared these with the J-band spectra
obtained at UKIRT.  We found that the flux calibration was accurate to
$\sim\!20\%$, with no obvious systematic offsets, and that the
differences in the power-law slope $\beta$ of the spectra were less
than $\sim\!0.1$.

Figure~\ref{fig:isabs} shows the observed flux distribution derived
from our spectroscopy, as well as from photometry in JHKL by Molnar
(\cite{moln:88}).  The figure also shows the flux distributions
corrected for three different amounts of interstellar extinction,
corresponding to $A_{\rm{}J}=4$, 5.5, and~7.  As above, we used the
reddening law of Mathis (\cite{math:90}).  For wavelengths longward of
0.8\,\micron, this extinction law is equivalent to
$A_\lambda\propto\lambda^{-\alpha}$, with $\alpha=1.7$.  For
comparison, we also show our I, J and K-band spectra of WR\,134 (WN6),
dereddened using $A_{\rm{}J}=0.69$ ($E_{B-V}=0.47$; Morris et al.\
\cite{morr&a:93}).  

From the figure, it is clear that the shape of the intrinsic flux
distribution of \CygX3 is very sensitive to the assumed amount of
interstellar reddening.  In order to obtain a flux distribution that
neither starts to fall at shorter wavelengths, nor becomes steeper
than the Rayleigh-Jeans tail of the Planck function, the value of
$A_{\rm{}J}$ should be larger than 4 and smaller than~7.  From more
detailed plots, we find that, in order to have $-0.3\la
\beta_{\rm{}\/I}\la2$ for the I-band spectra, one needs $5\la
A_{\rm{}J}\la6$.  [Here, $\beta=-0.3$ corresponds to optically thin
free-free emission.]  For this range, the power-law index as derived
from the longest wavelengths (i.e., K, L), varies from
$\beta_{K-L}=1.2$ to 1.6.  For $A_{\rm{}J}=5.5$, the whole flux
distribution longward of $0.8\,\mu$m is consistent with
$\beta_{\lambda>0.8}=1.4$.

An uncertainty in the above derivation is introduced by the
uncertainty of about 0.1 in the index $\alpha$ used in the extinction
law (Mathis \cite{math:90}).  We find that, to have $-0.3\la\beta_{\rm
\/I}\la2$ for $\alpha=1.8$, the range in $A_{\rm{}J}$ has to decreased
by $\sim\!0.1$, $\beta_{K-L}$ by $\sim\!0.2$, and
$\beta_{\lambda>0.8}$ by $\sim\!0.2$.  For $\alpha=1.6$, the range in
$A_{\rm{}J}$ has to be increased by $\sim\!0.3$, and $\beta_{K-L}$ by
$\sim\!0.1$.  For this value of $\alpha$, it is not possible to fit
the whole spectrum well with a single power-law, but to a reasonable
approximation $\beta_{\lambda>0.8}\simeq1.9$.

In summary, under the assumption that $-0.3\la\beta_{\rm{}I}\la2$, we
find that $A_{\rm{}J}=5.5\pm0.6$.  This value corresponds to a K-band
extinction of $A_{\rm K}=2.1\pm0.4$.  For $A_{\rm{}J}\simeq5.5$, the
flux distribution for $\lambda>0.8\,\micron$ has a power-law shape,
with $\beta_{\lambda>0.8}=1.4\pm0.3$.  This value is higher than
predicted by our model ($\beta=2/3$; see
Sect.~\ref{sec:modeldescription}), but not inconsistent with the
values found for normal Wolf-Rayet stars.  In the sample of 78
Wolf-Rayet stars studied by Morris et al.  (\cite{morr&a:93}), 6 have
$\beta\geq1.4$, and 20 have $\beta\geq1.1$.  Since we do not see a
strong variation in slope between the strong and the weak-line spectra
(Figs.~\ref{fig:isabs}, \ref{fig:iav} and \ref{fig:kav}), the large
slope we observe is unlikely to be related to the breakdown of the
model for high values of the temperature ratio $Q_T$.  Instead, it
more likely reflects the fact that, in general, the model is too
simple to provide a detailed description of a Wolf-Rayet wind.

\section{Discussion and conclusions\label{sec:bigdisc}}

The first infrared spectroscopic observations of \CygX3 (Paper~I)
showed the presence of Wolf-Rayet emission features in the infrared
spectrum of \CygX3, which indicates that a strong, dense, helium-rich
wind is present in the system, in which both the infrared lines and
continuum originate.  In follow-up observations, the emission lines
were much weaker, and they shifted in wavelength as a function of
orbital phase (Paper~II).  It was argued that this could be understood
in the context of the helium-star model, if the Wolf-Rayet wind is
sometimes strongly ionised and heated by the X~rays originating from
the compact object.  Furthermore, it was argued that if this were the
case, the observed modulation of the infrared continuum would follow
as a natural consequence.  In support of this argument, it was shown
that observed variations could be reproduced with a simple model of
the system, in which the wind has a two-temperature structure.

In this paper, we have presented a number of additional observations.
These observations confirm the prediction made in Paper~II that the
emission lines show wavelength shifts as a function of orbital phase
if they are weak, but not if they are strong.  We have described the
model used in Paper~II in detail, and presented model continuum light
curves and model line profiles for various combinations of the
parameters.  From these results, we have found that from the
assumption of a two-temperature wind it follows naturally not only
that the lines shift in wavelength, but also that the profiles at
maximum redshift are stronger and narrower than at maximum blueshift.

A problem we have encountered, is that the estimate of mass-loss rate
inferred from the infrared flux is an order of magnitude larger than
the estimate inferred from the rate of increase of the orbital period.
We have also found that the observed flux distribution is not
consistent with the model.  However, since similar deviations are
found for normal Wolf-Rayet stars, we believe that these discrepancies
reflect the simplifications made for our model, and that they do not
invalidate our main conclusions, viz., that the continuum and the
lines arise in the wind of the helium star, and that the modulation of
both lines and continuum is due to the wind being highly ionised by
the \Xray\ source, except in the helium star's shadow.

An open question that remains is what the underlying cause is of the
changes in state, in the infrared as well as in \Xray\ and radio.
Watanabe et al.\ (\cite{wata&a:94}) have found that the radio and
\Xray\ states are correlated: large radio outbursts occur when the
\Xray\ source is in its high state (high flux, soft spectrum), but not
when it is in its low state (low flux, hard spectrum).  Kitamoto et
al.\ (\cite{kita&a:94}) determined that the radio/\Xray\ source was in
its high state in June 1991, when the infrared spectra showed strong
lines, and in its low state in May 1992, when the lines were weak.
They suggested that the state at all wavelengths was a function of the
mass-loss rate of the Wolf-Rayet star only.  When high, the infrared
continuum would be stronger and the accretion rate would be high,
leading to a strong \Xray\ source and radio outbursts (the latter are
presumably due to the ejection of relativistic jets; e.g., Geldzahler
et al.\ \cite{geld&a:83}).  The wind would be dense enough to be
optically thick to low-energy X~rays, leading to a low state of
ionisation.  When the mass-loss rate was lower, the infrared continuum
flux and the accretion rate would be low, leading to \Xray\ low state
and the absence of radio outbursts.  They suggested that the wind
would become optically thin to low-energy X~rays, so that it could be
ionised despite the lower \Xray\ luminosity.

At present, the hypothesis of a variable wind mass-loss rate as the
underlying cause of the changes in state is consistent with the
observations.  Other Wolf-Rayet stars of the WN subclass, however, do
not show evidence for large changes in mass-loss rate, showing, e.g.,
constant flux to within a few percent on all observed time scales
(Moffat \& Robert \cite{moffr:91}).  It might be that the changes in
accretion rate and optical depth that are required are due to
processes near the \Xray\ source and its accretion disk, perhaps
similar to those underlying the changes in state observed in
black-hole candidates like \CygX1 (e.g., Oda \cite{oda:77}).

One possibly major problem that has been addressed only briefly so far
(in Paper~I) is that the radii of Wolf-Rayet stars derived from model
fits to Wolf-Rayet spectra (e.g., Schmutz et al.\ \cite{schm&a:89}),
are much larger than the radius of the Roche lobe in \CygX3 (Paper~I;
Conti \cite{cont:92}).  Schmutz (\cite{schm:93}) has tried to model
the 1991 I and K-band spectra of \CygX3 using the Wolf-Rayet wind
model described by Hamann \& Schmutz (\cite{hamas:87}) and Wessolowski
et al.\ (\cite{wess&a:88}).  He found that the stellar parameters were
typical for a Wolf-Rayet star of spectral type WN7, but that given the
absolute K magnitude, the photospheric radius was larger than the
orbital separation by a factor~3.  In order to resolve this
discrepancy, he suggested that the distance to \CygX3 was much smaller
than 10\,kpc, so that the luminosity, and thus the mass and radius,
were much smaller.  He suggested that the 21\,cm absorption features,
on which the 10\,kpc distance is based, might be due to circumstellar
shells.

Although the suggestion of a lower distance is tempting in that it
would also resolve the discrepancy between the different estimates of
the mass-loss rate, we believe that it is unlikely that there are
circumstellar shells at such velocities and with such strengths that
they produce a 21\,cm absorption profile which corresponds closely
with the combined emission profile of the Local, Perseus and Outer
arms (e.g., Dickey \cite{dick:83}).  Instead, we believe it more
likely that the estimate of the radius derived from the
model-atmosphere fits is wrong.  We stress that these estimates are
based on an {\em{}assumed} velocity law, which is used to extrapolate
inward from the regions where the continuum and lines are formed, to
the so-called `zero-velocity' radius.  Usually, for this velocity law,
a $\beta$-law is used ($v=v_\infty(1-R_*/r)^\beta$).  This velocity
law has been found to be a reasonable approximation for O stars, but
there is no reason to assume it is valid for Wolf-Rayet stars as well,
especially close to the star (for more thorough discussions, we refer
to Kudritzki \& Hummer \cite{kudrh:90}; Hillier \cite{hill:95};
Schmutz \cite{schm:95}).

An observational indication that the radii of Wolf-Rayet stars are
actually quite small comes from studies of Wolf-Rayet stars in
eclipsing binaries.  For instance, Cherepashchuk et al.\
(\cite{cher&a:84}) found that the eclipse light curve of V444~Cyg
indicated a core radius (electron-scattering optical depth unity) of
only 2.8\,\Rsun.  Hamann \& Schwartz (\cite{hamas:92}) argue that this
estimate is not unique, and that the radius of the Wolf-Rayet star
could be much larger.  From polarisation measurements, however,
St-Louis et al.\ (\cite{stlo&a:93}) find that radii larger than
$\sim\!4\,\Rsun$ are excluded.  A similar conclusion is reached by
Cherepashcuk et al.\ (\cite{cher&a:95}) based on a spectroscopic
estimate of the luminosity ratio.  Using this and other observational
evidence, Moffat \& Marchenko (\cite{moffm:96}) conclude that the
radii of Wolf-Rayet stars are not unlike the radii expected from
stellar model calculations (e.g., Langer \cite{lang:89}).  For such
radii, the helium star would fit well inside the Roche lobe in \CygX3
(Paper~I).

We conclude from our data that the idea that the companion of \CygX3
is a massive Wolf-Rayet star is entirely plausible.  It seems to us
that \CygX3 is a very interesting system, in which a Wolf-Rayet wind
can be probed through the combined effects of occultation and variable
ionisation by X~rays.

\begin{acknowledgements} 
We thank Rob Fender and Werner Schmutz for stimulating discussions,
George Herbig for identifying the interstellar absorption feature in
the I-band spectra, Malcolm Coe for providing us with infrared
photometry, Rob Koopman for careful reading of the manuscript, and
Jean-Marie Vreux for his thorough review.  This research made use of
the SIMBAD database, operated at CDS, Strasbourg, France.  It was
supported in part by the Netherlands Organisation for Scientific
Research NWO (MvdK, MHvK), by the Leids Kerkhoven-Bosscha Fonds
(MHvK), and by NASA through a Hubble fellowship (MHvK) awarded by
STScI.
\end{acknowledgements}

\end{document}